\documentclass[useAMS,usenatbib,referee]{biom}
\def\bSig\mathbf{\Sigma}

\title[Adaptive Bayesian Sum of Trees Model for Covariate Dependent Spectral Analysis]{Adaptive Bayesian Sum of Trees Model for Covariate Dependent Spectral Analysis}

\author{Yakun Wang$^{1}$, 
Zeda Li$^{2}$, and 
Scott A. Bruce$^{1,*}$\email{sbruce7@gmu.edu} \\
$^{1}$Department of Statistics, George Mason University, Fairfax, Virginia, U.S.A. \\
$^{2}$Paul H. Chook Department of Information System and Statistics, \\Baruch College, The City University of New York, New York, New York, U.S.A.
}

\usepackage{amsmath}
\usepackage{graphicx}
\usepackage{subfigure}
\usepackage{amsfonts}

\usepackage{natbib}
\usepackage[hidelinks]{hyperref}
\usepackage{booktabs}
\usepackage{url} 
\usepackage{todonotes}

\begin{document}





\pagerange{\pageref{firstpage}--\pageref{lastpage}} 




\pagerange{} 
\volume{}
\pubyear{}
\artmonth{}

\doi{}

\label{firstpage}


\begin{abstract}
This article introduces a flexible and adaptive nonparametric method for estimating the association between multiple covariates and power spectra of multiple time series. The proposed approach uses a Bayesian sum of trees model to capture complex dependencies and interactions between covariates and the power spectrum, which are often observed in studies of biomedical time series.  Local power spectra corresponding to terminal nodes within trees are estimated nonparametrically using Bayesian penalized linear splines.  The trees are considered to be random and fit using a Bayesian backfitting Markov chain Monte Carlo (MCMC) algorithm that sequentially considers tree modifications via reversible-jump MCMC techniques. For high-dimensional covariates, a sparsity-inducing Dirichlet hyperprior on tree splitting proportions is considered, which provides sparse estimation of covariate effects and efficient variable selection. By averaging over the posterior distribution of trees, the proposed method can recover both smooth and abrupt changes in the power spectrum across multiple covariates.  Empirical performance is evaluated via simulations to demonstrate the proposed method's ability to accurately recover complex relationships and interactions.  The proposed methodology is used to study gait maturation in young children by evaluating age-related changes in power spectra of stride interval time series in the presence of other covariates.
\end{abstract}

%

\begin{keywords}
Bayesian backfitting; Gait variability; Multiple time series; Reversible jump Markov chain Monte Carlo; Spectrum analysis; Whittle likelihood.
\end{keywords}


\maketitle


%

\section{Introduction}
\label{s:intro}

The frequency-domain properties of time series have often been found to contain valuable information. For example, frequency-domain analysis of biomedical time series, such as gait variability, heart rate variability (HRV), and electroencephalography (EEG), provides interpretable information about underlying physiological processes \citep{HausdorffJ.M1999Mogd,Halletal2004,Klimesch1999}. In many studies, biomedical time series are collected from multiple participants in conjunction with multiple covariates to explore connections between prominent oscillatory patterns in the time series and various clinical and behavioral outcomes.  These relationships are often complex and highly interactive.
As a result, a flexible, adaptive method that can estimate the association between power spectra and multiple covariates is needed to better understand the complex relationships between physiological processes and important measures of health and functioning.

\begin{figure}
  \centering
    \includegraphics[width=.8\linewidth]{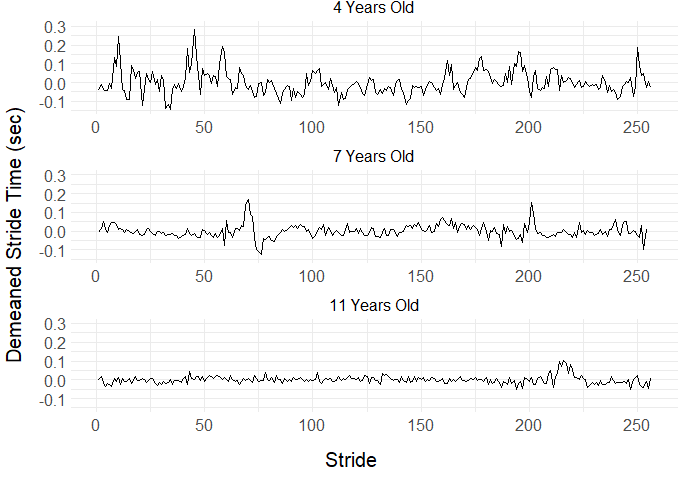}
   \caption{Demeaned stride interval time series for three participants in the gait maturation study ages 4, 7, and 11 years old.}
  \label{fig:Timeseries}
\end{figure}

A prime example and motivating application for this article comes from a study of maturation in gait dynamics in young children \citep{HausdorffJ.M1999Mogd}.  Immature gait in very young children results in unsteady walking patterns and frequent falls \citep{article,VanderLindenDarlW1996SAWM}.  While gait is relatively mature by age 3, neuromuscular control continues to develop well beyond this age \citep{VanderLindenDarlW1996SAWM,PreisSabine2008Gabm}.  Accordingly, it is of interest to assess if gait dynamics continue to become more steady and regular beyond age 3, in conjunction with improving neuromuscular control.  To assess gait variability and posture control in younger children, stride interval time series consisting of stride times during normal walking were observed in fifty children between the ages of 3 and 14 \citep{HausdorffJ.M1999Mogd}.  For illustration, Figure \ref{fig:Timeseries} displays demeaned stride interval time series for three participants age 4, 7, and 11 years old.  In addition to age, other covariates were also collected that may influence gait, such as gait speed and gender.  In quantifying the association between age and the power spectra of stride interval time series, we seek to better understand the maturation of gait dynamics and variability associated with developing neuromuscular control with age in the presence of other related covariates. 

In the time series literature, spectral analysis of multiple time series has received much attention in recent years. To quantify the association between a {\it single} covariate and power spectra, \cite{FiecasMark2017MtEo} and \cite{krafty2017} propose methods that can capture a smooth covariate effect on the power spectrum. \cite{BruceScottA2018CaBs} proposes an adaptive Bayesian method that can capture both smooth and abrupt changes in power spectra across a covariate. \cite{LiBruce2020MultiCABS} adapts the method of \cite{BruceScottA2018CaBs}  for covariate-dependent spectral analysis of replicated multivariate time series. However, these methods are not readily extendable to incorporate {\it multiple} covariates, which hinders their applicability to many important studies. Existing methods that can account for multiple covariates are either parametric \citep{diggle1997spectral} or semi-parametric \citep{iannaccone2001semiparametric,QinLi2009ATFM,stoffer2010smoothing,krafty2011,chauandsachs2016}. These approaches characterize covariate effects via design matrices within a linear modeling framework, and thus can not immediately accommodate complex dependencies and interactions among covariates and power spectra. One exception is the approach proposed by \cite{BertolacciMichael2019ACDS}, which introduces a mixture modeling approach with covariate-dependent mixture weights to account for complex covariate effects. However, a thin-plate Gaussian process prior is imposed on mixture weights, which is both smooth and stationary, and may not be appropriate for modeling abrupt spectral dynamics. Moreover, this method does not provide a means for variable selection when a large number of covariates are observed. The goal of this article is to introduce a flexible method that can capture both smooth and abrupt changes in power spectra across multiple covariates, without loss of interpretation, and simultaneously provide a tool for variable selection.  


To capture complex smooth, abrupt, and interaction effects of covariates on power spectra in a parsimonious manner, we propose a tree-based covariate partitioning framework. Tree-based models are not new and have become extremely popular in recent years \citep{BreimanLeo2001RF,ChipmanHughA2010BBar}. For example, \cite{ChipmanHughA2010BBar} propose a Bayesian additive regression tree (BART) model which can flexibly model complex covariate effects and interactions and demonstrates outstanding predictive performance \citep{Chipman2013}. Accordingly, BART has been widely applied in many different scientific domains for various types of outcomes \citep{WaldmannPatrik2016GpuB,BlattenbergerGail2017TAFM,vanderMerweSean2018TSAo} including smooth functional response variables \citep{starling2018bart}.  

In this article, a sum of trees model for the covariate-dependent power spectrum is introduced to simultaneously partition multiple covariates parsimoniously. A penalized linear spline model is used for local spectrum estimation within terminal nodes of the trees. The framework is formulated in a fully Bayesian setting where the trees are random and fit using an iterative Bayesian backfitting Markov chain Monte Carlo (MCMC) procedure and reversible-jump techniques \citep{green95} to evaluate various tree modifications. 

The proposed methodology expands the scope of covariate dependent power spectra that can be accurately recovered in three meaningful ways. First, the flexible sum of trees model can recover complex nonlinear relationships and interaction effects without assuming a particular form of the relationship a priori. Second, by averaging over the posterior distribution of trees, the proposed method can recover both smooth and abrupt covariate effects on the power spectrum. Third, the proposed method can automatically accommodate mixed-type covariates (nominal, ordinal, discrete, continuous) through the underlying tree structures, as well as high-dimensional covariates by placing a sparsity-inducing Dirichlet hyperprior on the splitting proportions of the regression tree prior \citep{LineroAntonioR2018BRTf} for sparse estimation of covariate effects and variable selection.

The rest of the paper is organized as follows. Section \ref{sec:Covariate-Dependen} provides a definition for the covariate-dependent power spectrum. The tree-based modeling framework for the power spectrum is introduced in Section \ref{sec:Approximately stationary}, and Section \ref{sec:method} proposes the adaptive Bayesian sum of trees model and MCMC sampling scheme. Simulation results for various covariate effects (e.g. linear, nonlinear, smooth, abrupt, high-dimensional) and interaction effects are provided in Section \ref{sec:sim}. Section \ref{sec:app} presents the application to the study of gait maturation in young children. Conclusions and future directions of this work are covered in Section \ref{sec:end}.

\section{Covariate-Dependent Power Spectrum}
\label{sec:Covariate-Dependen}
We consider modeling a collection of stationary time series $X_{\ell t}$ of length $t=1,\ldots,T$ and $P$-dimensional covariate vectors $\boldsymbol{\omega}_{\ell}=(\omega_{1 \ell},\ldots,\omega_{P \ell})'$ 
for $p=1,\ldots,P$ mixed-type covariates and $\ell =1,\ldots,L$ independent subjects.  
To evaluate the effect of multiple covariates on the power spectrum, an extension of the Cram\'{e}r representation \citep{cramer42}
is considered, which allows the power spectrum to vary across frequencies and covariates
\begin{equation*}\label{covaraite_cramer}
X_{\ell t} =\int^{1/2}_{-1/2} A(\boldsymbol{\omega}_{\ell},\nu)\exp(2\pi it\nu)dZ_{\ell}(\nu),
\end{equation*}
where $A(\boldsymbol{\omega}_{\ell},\nu)$ is a Hermitian and periodic complex-valued function of frequency $\nu \in \mathbb{R}$ and covariates $\boldsymbol{\omega_{\ell}}$ such that $A(\boldsymbol{\omega}_{\ell}, \nu)=\overline{A(\boldsymbol{\omega}_{\ell}, -\nu)}$,  $A(\boldsymbol{\omega}_{\ell}, \nu) = A(\boldsymbol{\omega}_{\ell}, \nu + 1)$, and $A(\boldsymbol{\omega}_{\ell},\nu)$ is square integrable over frequencies $[-1/2,1/2]$. $Z_{\ell}$ are zero-mean mutually independent and identically distributed orthogonal processes with unit variance. Regularity conditions on the distribution of $Z_{\ell}$ must also be assumed to ensure subsequently introduced estimators are well-behaved; we assume cumulants of $dZ_{\ell}$ exist and are bounded for all orders \citep{brillinger2001}. 

The covariate-dependent power spectrum is then defined as $f(\boldsymbol{\omega},\nu)=|A(\boldsymbol{\omega},\nu)|^{2}$ and can be interpreted as the contribution to the variance at frequency $\nu$ given covariate values $\boldsymbol{\omega}$. We assume that $A$, and subsequently the spectrum $f$, are continuous functions of frequency $\nu$, but can have a finite number of discontinuities as functions of covariates $\boldsymbol{\omega}$. This flexibility allows for modeling abrupt changes over the covariate space and differs from models assuming entirely smooth changes across covariates \citep{FiecasMark2017MtEo,krafty2017}.

\section{Tree-Based Modeling of the Power Spectrum}
\label{sec:Approximately stationary}
\subsection{Tree-Based Piecewise Stationary Approximation}

The covariate-dependent power spectrum introduced in Section \ref{sec:Covariate-Dependen} can be approximated by piecewise stationary processes via partitioning the covariate space into approximately stationary blocks. Piecewise stationary approximations of the power spectrum have been well-studied within the context of nonstationary time series analysis \citep{Adak1998, rosen2012}, where time series are divided into approximately stationary intervals for time-dependent spectral analysis.  \cite{BruceScottA2018CaBs} introduced a time- and covariate-based piecewise stationary approximation for time- and covariate-dependent spectral analysis using a two-dimensional grid.  

However, directly extending a two-dimensional grid to higher dimensions in order to accommodate multiple covariates can easily lead to over-parameterization and undue computational complexity. Tree-based approaches represent a more flexible and parsimonious alternative for partitioning multiple covariates. 
A grid-based partition tends to produce a finer partition than is necessary, since partition points for each covariate do not depend on the other covariate. This results in less efficient information sharing across series with similar covariates and less accurate estimation. Conversely, if the true partition does have a grid structure, a tree-based model can still well-approximate the partition and is thus preferable.

    

For tree-based piecewise stationary approximation, each terminal node is defined through a collection of splitting rules corresponding to the tree structure and represents an approximately stationary region of the covariate space. Suppose a tree $U$ has $B$ terminal nodes.  The corresponding piecewise stationary approximation is
\begin{equation*}\label{eq:picewise}
\begin{split}
X_{\ell t} & \approx  \sum_{b=1}^B \delta(\boldsymbol{\omega}_{\ell};U,b)X_t^{(b)} =\int^{1/2}_{-1/2} \sum_{b=1}^B \delta(\boldsymbol{\omega}_{\ell};U,b)A_b(\nu)\exp(2\pi it\nu)dZ_{\ell}(\nu),
\end{split}
\end{equation*}
where $X_t^{(b)}$ is stationary process with power spectrum $f_b(\nu)=|A_b(\nu)|^{2}$ corresponding to the $b$th terminal node, and $\delta$ is a function that identifies terminal node membership for each observation based on covariates such that $\delta(\boldsymbol{\omega}_{\ell};U,b)=1$ if the $\ell$th observation falls into the $b$th terminal node and $\delta(\boldsymbol{\omega}_{\ell};U,b)=0$ otherwise.

\subsection{Local Power Spectrum Estimation}
\label{sec:localest}
We now introduce an estimator for local power spectra within terminal nodes of the tree. Let $N= \lfloor T/2 \rfloor - 1$ and $\nu_k = k/T$ for $k=1,\ldots,N$ be the Fourier frequencies.  The periodogram estimator of the power spectrum for the $\ell$th time series is
$I_{\ell}(\nu_k) =\frac{1}{T}\left|\sum_{t=1}^{T}X_{\ell t}\exp(-2\pi i\nu_k t)\right|^{2}.$
The Whittle likelihood \citep{Whittle1952}, derived from the large sample distribution of the periodogram, can then be used to approximate the overall likelihood for the tree as a product of individual likelihoods, assuming $T$ is sufficiently large,
\begin{multline}\label{eq:Whittle}
L(I_{1},\ldots,I_{\ell}|f_{1},\ldots,f_{B}) \approx \\ \prod^{L}_{\ell=1}\prod^{B}_{b=1}(2\pi)^{-N/2} \prod^{N}_{k=1}\exp\{-\delta(\boldsymbol{\omega}_{\ell};U,b)[\log f_{b}(\nu_{k})+\exp(\log I_{\ell}(\nu_{k}) - \log f_{b}(\nu_{k}))]\}.
\end{multline}

Log power spectra within each terminal node $\log f_{b}(\nu)$ are modeled using a Bayesian penalized linear spline model \citep{rosen2012}
\begin{equation}\label{eq:splinemodel}
        \log f_{b}(\nu)\approx \alpha_{b}+\sum_{s=1}^{S}\beta_{s}^{(b)}\cos(2\pi s\nu),
\end{equation}
where the functions $\cos(2\pi s \nu)$ are the Demmler–Reinsch basis functions for periodic even splines observed on an evenly spaced grid (i.e. the Fourier frequencies) \citep[Section 3]{SchwarzKatsiaryna2016Auff}.  Only the first $S < N$ basis functions are used to provide a low-rank approximation to the full linear smoothing spline \citep{eubank1999}.  
In order to achieve good computational efficiency without sacrificing estimation accuracy \citep{krafty2017}, $S=7$ basis functions are used for subsequent simulations and real data analyses, which provide good empirical performance. 
Gaussian priors are assumed such that $\alpha_{b}\sim N(0,\sigma_{\alpha}^{2})$ where $\sigma_{\alpha}^{2}$ is a large constant value, and $\boldsymbol{\beta}^{(b)}=(\beta_{1}^{(b)},\ldots,\beta_{S}^{(b)})' \sim N(0,\tau_{b}^{2}\boldsymbol{D}_{S})$, where $\boldsymbol{D_{S}}=\text{diag}(\{\sqrt{2}\pi s\}^{-2})$. $\tau_b^2$ is a smoothing parameter that controls the roughness of the log spectrum.  The scaling for the smoothing parameter, $\{\sqrt{2}\pi s\}^{-2}$, provides regularization of the integrated squared first derivative of the log power spectrum \citep{li2019adaptive}. A half-t prior is placed on $\tau_{b}$ \citep{gelman2006} to complete the Bayesian model specification. We follow \cite{wand2011} and express the half-t distribution as a scale mixture of inverse gamma distributions for efficient sampling. A two-step MCMC sampling scheme for $\alpha_{b}$, $\boldsymbol{\beta}^{(b)}$, and $\tau_{b}$ following \cite{rosen2012} is presented below.     
\begin{enumerate}
\item Let $\boldsymbol{Z}_b$ be a $N\times S$ matrix of basis functions for $b$th terminal node such that $\{Z_b\}_{k,s} = \cos(2\pi s \nu_k)$.
Given $\tau_{b}$, basis functions $\boldsymbol{Z}_b$, and periodogram ordinates $\boldsymbol{I}_{\ell}(\nu)=\{I_{\ell}(\nu_{1}),$ \\ $I_{\ell}(\nu_{2}),\ldots,I_{\ell}(\nu_{N})\}$ for $l=1,\ldots,L$, $\alpha_{b}$ and $\boldsymbol{\beta}^{(b)}$ are sampled jointly in a Metropolis-Hastings (M-H) step from 
\begin{equation}\label{eq:Gibbsbeta}
    \begin{split}
        p(\alpha_{b},\boldsymbol{\beta}^{(b)}|\tau_{b}^{2},\boldsymbol{I}_{\ell}(\nu),\boldsymbol{Z}_b) & \propto \exp\Big\{-\sum_{\ell=1}^{L}\sum_{k=1}^{N}[\alpha_{b}+\mathbf{z}_{bk}'\boldsymbol{\beta}^{(b)}+\exp\big(\log I_{\ell}(\nu_{k})\\
         & -\alpha_{b}- \mathbf{z}_{bk}'\boldsymbol{\beta}^{(b)}\big)]-\frac{\alpha_{b}}{2\sigma_{a}^{2}}-\frac{1}{2\tau_{b}^{2}}\boldsymbol{\beta}^{(b)'}\boldsymbol{D}_{S}^{-1}\boldsymbol{\beta}^{(b)}\Big\}.
    \end{split}
 \end{equation}

 \item By representing the half-t prior as a scale mixture of inverse gamma distributions \citep{wand2011}, we obtain draws of $\tau_{b}$ from its full conditional distribution by sampling from
 \begin{equation}\label{posterior_tau1}
         (a_{b}|\tau_{b}^{2})\sim \text{IG}\Big(\frac{\xi_{\tau}+1}{2},\frac{\xi_{\tau}}{\tau_{b}^{2}}+\frac{1}{A_{\tau}^{2}}\Big),
 \end{equation}
 and
 \begin{equation}\label{posterior_tau2}
         (\tau_{b}^{2}|a_{b},\boldsymbol{\beta}^{(b)})\sim \text{IG}\Big(\frac{\xi_{\tau}+S+1}{2},\frac{\boldsymbol{\beta}^{(b)'}\boldsymbol{\beta}^{(b)}}{2}+\frac{\xi_{\tau}}{a_{b}}\Big)
 \end{equation}
where $a_b$ is a latent variable, and $\xi_{\tau}$ and $A_{\tau}^2$ are fixed hyperparameters of the inverse gamma distribution.
 \end{enumerate}


\section{Adaptive Bayesian Sum of Trees Model}
\label{sec:method}
\subsection{Sum of Trees Model}
Poor mixing of single tree models has been noted in many applications such that the MCMC algorithm becomes stuck in subsets of the covariate space representing local optima and cannot efficiently traverse the entire parameter space \citep{WuYuhong2007BCPS}.  This can happen when single tree models grow very large in an effort to approximate more complex relationships, thus restricting possible modifications due to low sample size and an abundance of other splits. We follow \cite{ChipmanHughA2010BBar} in developing a Bayesian sum of trees formulation to relieve this problem by constructing many shallow trees as ``weak learners" for estimation. 

Let $M$ be the number of trees. A sum of trees model for the log power spectrum is then constructed as
\begin{equation*}\label{eq:modelspectrum}
\log f(\boldsymbol{\omega},\nu) \approx \sum_{j=1}^{M}\sum_{b=1}^{B_j} \delta(\boldsymbol{\omega};U_j,b)\log f_{bj}(\nu),
\end{equation*}
where $U_j$ represents the $j$th tree that has $B_j$ terminal nodes for $j=1,\ldots,M$.  Model specification for local power spectra $\log f_{bj}(\nu)$ within each tree then follows directly from the specification for the single tree model introduced in Section \ref{sec:localest}.

\subsection{Prior Specification}
Let $\Phi_j = \{\log f_{1j}(\nu),\ldots,\log f_{B_j j}(\nu)\}$ be the collection of log power spectra across terminal nodes for the $j$th tree.  To complete the Bayesian model specification, priors are imposed on $U_j$ and $\Phi_j$ in order to allow the trees to be random and fit from the data. Assuming independence across terminal node parameters and trees a priori, priors can be specified as

\begin{equation*}
p((U_{1},\Phi_{1}),...,(U_{M},\Phi_{M})) =\prod_{j}p(U_{j},\Phi_{j})=\prod_{j}p(\Phi_{j}|U_{j})p(U_{j}),
\end{equation*}
where
\begin{equation*}
p(\Phi_{j}|U_{j})
=\prod_{b}p(\log f_{bj}(\nu)|U_{j}).
\end{equation*}


The priors for $\Phi_j|U_j$ then correspond to the priors of $\alpha_b$, $\boldsymbol{\beta}^{(b)}$ and $\tau_b$ for the local power spectrum estimator introduced in Section \ref{sec:localest}. For the priors on the tree structure $U_j$, three probabilities need to be considered. 
\begin{enumerate}
    \item The probability of a node to be split is defined as $\Pr(\text{SPLIT})=\gamma(1+d)^{-\theta}$, where $\gamma\in(0,1), \ \theta\in[0,\infty)$ and $d=0,1,\ldots$ is the depth of a given node. This prior is a regularization of the tree depth to encourage each tree to be shallow. We set $\gamma=0.95$ and $\theta=2$ following \citet{ChipmanHughA2010BBar}. \cite{RockovaVeronika2018OTfB} propose a minor modification $\Pr(\text{SPLIT}) \propto \gamma^{d}$ for some $0 \le \gamma < 1/2$ to achieve the optimal posterior convergence rate, which can also be adopted within the proposed framework.
    
    \item The probability of selecting the $p$th covariate for splitting is denoted as $s_p$ for $p=1,\ldots,P$. The proposed model allows for two possible prior specifications: a uniform prior, $s_p = P^{-1}$, such that all covariates have the same probability to be selected, and a sparsity-inducing Dirichlet prior         $(s_{1},\ldots,s_{P})\sim \mathcal{D}\Big(\frac{\sigma}{P},\ldots,\frac{\sigma}{P}\Big)$ \citep{LineroAntonioR2018BRTf}.  For the Dirichlet prior, $\sigma$ determines the degree of sparsity and \cite{LineroAntonioR2018BRTf} offer multiple approaches for modeling this parameter.  We set $\sigma=1$ as suggested by \cite{LineroAntonioR2018BRTf} for computational convenience. 

  \item The probability of selecting a particular cutpoint for a given covariate is uniform across all cutpoints. A uniform prior is also desirable as it is invariant for monotone transformations of the covariate \citep{ChipmanHughA2010BBar}. For continuous, discrete, and ordinal covariates, cutpoints are selected from a fixed number of evenly spaced points over the range of possible values.  For categorical covariates without an intrinsic ordering, a cutpoint represents a particular mapping of categories to the left and right child nodes created by the split.  A categorical variable with $q$ categories then has $2^q-2$ cutpoints that can be selected.
  
  
\end{enumerate}

\subsection{Bayesian Backfitting MCMC}

It is important to note that each tree is capturing particular features of the covariate-dependent power spectrum and also depends on the features captured by other trees. While this provides considerable flexibility and adaptive estimation, it presents significant computational challenges in estimating the trees.  Following \cite{ChipmanHughA2010BBar}, we develop a Bayesian backfitting MCMC algorithm for proposing and evaluating modifications to each tree sequentially.

For the $j$th tree, the posterior distribution $p\big((U_{1},\Phi_{1}),...,(U_{M},\Phi_{M})|\boldsymbol{I}_1(\nu_k),\ldots,\boldsymbol{I}_L(\nu_k)\big)$ can be sampled through $M$ successive draws from 
$p\big((U_{j},\Phi_{j})|U_{-j},\Phi_{-j},\boldsymbol{I}_1(\nu_k),\ldots,\boldsymbol{I}_L(\nu_k)\big)$ where $U_{-j}$ is the set of all trees except $U_{j}$, and $\Phi_{-j}$ is defined similarly. Observe that the conditional distribution $p\big((U_{j},\Phi_{j})|U_{-j},\Phi_{-j},\boldsymbol{I}_1(\nu_k),\ldots,\boldsymbol{I}_L(\nu_k)\big)$ depends on $\big(U_{-j},\Phi_{-j},\boldsymbol{I}_1(\nu_k),$\\$\boldsymbol{I}_2(\nu_k),\ldots,\boldsymbol{I}_L(\nu_k)\big)$ only through
\begin{equation*}
\mathbf{R}_{\ell j}(\nu_k)  = \log\boldsymbol{I}_{\ell}(\nu_k)-\sum_{i\neq j}\sum_{b=1}^{B_j} \delta(\boldsymbol{\omega};U_i,b)\log f_{bi}(\nu),
\end{equation*}
where $\mathbf{R}_{\ell j}(\nu_k)$ is the residual of the log periodogram after removing the fit from the sum of trees across all trees except for the $j$th tree. Therefore, drawing from the posterior distribution equates to $M$ successive draws from $p\big((U_{j},\Phi_{j})|\mathbf{R}_{1 j}(\nu_k),\ldots,\mathbf{R}_{L j}(\nu_k)\big)$.  Hence, the local power spectrum estimator for the $j$th tree is then fit by replacing $\log \boldsymbol{I}_{\ell}(\nu_k)$ with $\mathbf{R}_{\ell j}(\nu_k)$ in Equation \eqref{eq:Whittle} and Equation \eqref{eq:Gibbsbeta}.

\subsection{Reversible-jump MCMC Sampling}
\label{sec:RJMCMC}
To sample new tree structures, a reversible-jump MCMC (RJMCMC) \citep{green95} procedure is developed to jointly propose and evaluate new draws of $U$ and $\Phi$ for each tree. The proposed modification takes the form of one of three possible moves: BIRTH, DEATH and CHANGE, with probabilities 0.25, 0.25 and 0.5 respectively. The BIRTH move grows the tree by splitting a terminal node into two child nodes, the DEATH move prunes the tree by dropping two terminal child nodes belonging to the same internal node, and the CHANGE move modifies the variable and cut point associated with an internal node with two terminal child nodes. New model parameters for the proposed tree modification are drawn, and the proposed tree modification is then accepted or rejected using a Metropolis–Hastings (M-H) step. Each tree is considered in turn for updating within each iteration. Draws using this RJMCMC sampling scheme are repeated and averaged over post burn-in draws to obtain the final estimator.  Technical details for the RJMCMC sampling scheme are available in Web Appendix A. 


\section{Simulation Studies}
\label{sec:sim}

We consider three simulation settings representing abrupt and smoothly varying dynamics with complex covariate effects and interactions in order to demonstrate strong finite-sample estimation accuracy, as well as the ability to adapt to sparse covariate effects and conduct variable selection.

\subsection{Settings}
\label{sec:sim_setting}
\begin{enumerate}
    \item Abrupt+Smooth: Let $\boldsymbol{\omega} = (\omega_1,\omega_2)$ where $\omega_1,\omega_2 \stackrel{i.i.d.}{\sim} U(0,1)$. An AR(1) process for the $\ell$th time series is specified as
    $
        x_{\ell t}=\phi_{\ell}x_{\ell t-1}+\epsilon_{\ell t},  \; \epsilon_{\ell t}\sim N(0,1),
   $ where $\phi_{\ell}=-0.7+1.4\omega_{2}$ when $0\leq\omega_{1}< 0.5$ and $\phi_{\ell}=0.9-1.8\omega_{2}$, when $0.5\leq\omega_{1}\leq 1$. 
    
    \item AR-Friedman: Let $\boldsymbol{\omega} = (\omega_1,\ldots,\omega_5)$ where $\omega_1,\ldots,\omega_5 \stackrel{i.i.d.}{\sim} U(0,1)$. An AR(1) process for the $\ell$th time series is specified as
   $
        x_{\ell t}=\phi_{\ell}x_{\ell t-1}+\epsilon_{\ell t},  \; \epsilon_{\ell t}\sim N(0,1),
    $ where $  \phi_{\ell} =0.5\sin(\pi \omega_{\ell 1}\omega_{\ell 2})-(\omega_{\ell 3}-0.5)^{2}+0.35 \mathrm{ sign}( \omega_{\ell 4}-0.5)-0.15\omega_{\ell 5}$.

    \item Adjusted-AdaptSPEC-X: Let $\boldsymbol{\omega}=(\omega_{1},\omega_{2})$, where $\omega_{1},\omega_{2} \stackrel{i.i.d.}{\sim} U(0,1)$. Each covariate vector $\boldsymbol{\omega}$ is mapped to a latent variable $z_{\ell}\in\{1,2,3,4\}$. Figure \ref{fig:AdaptX_scatter} shows the mapping from $\boldsymbol{\omega}$ to $z$. An AR(2) process is then specified as
    $
    x_{\ell t}  =\phi_{z_{\ell}1}x_{\ell t-1}+\phi_{z_{\ell}2}x_{\ell t-2}+\epsilon_{\ell t}, \; \epsilon_{\ell t}\sim N(0,1),
    $ where the coefficients for the four latent region are defined as follows. If $z_{\ell}=1$, $(\phi_{z_{\ell}1}, \phi_{z_{\ell}2})=(1.5,-0.75)$; if $z_{\ell}=2$, $(\phi_{z_{\ell}1}, \phi_{z_{\ell}2})=(-0.8,0)$; if $z_{\ell}=3$, $(\phi_{z_{\ell}1}, \phi_{z_{\ell}2})=(-1.5,-0.75)$; if $z_{\ell}=4$, $(\phi_{z_{\ell}1}, \phi_{z_{\ell}2})=(0.2,0)$.

\end{enumerate}


\begin{figure}
    \centering
    \subfigure[Latent variable mapping \label{fig:AdaptX_scatter}]{\includegraphics[width=0.47\textwidth]{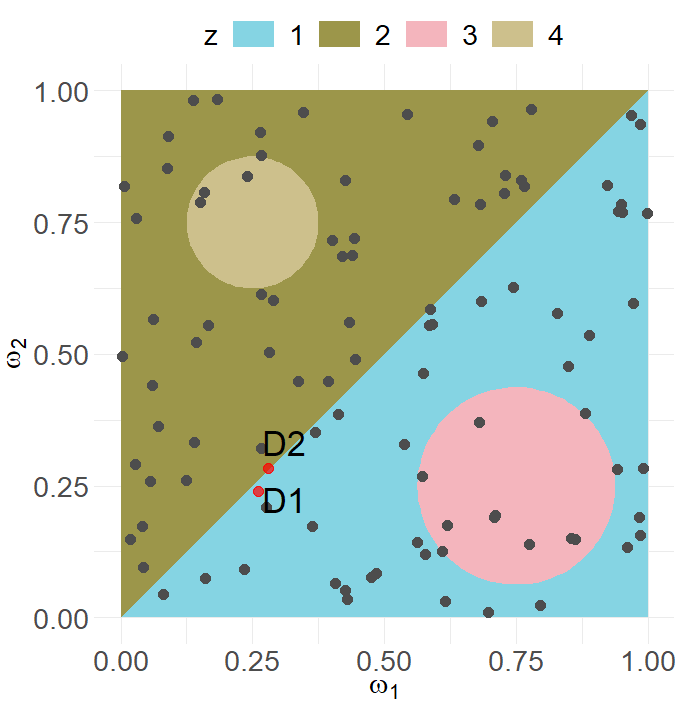}} 
    \subfigure[Model estimation \label{fig:Edgepoints_Comparision}]{\includegraphics[width=0.3\textwidth]{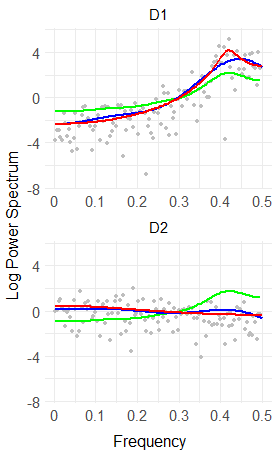}} 
    \caption{(a) presents the mapping of covariate values $\omega_1$ and $\omega_2$ to latent variable values $z$ for the Adjusted-AdaptSPEC-X simulation setting. D1 and D2 denote two simulated time series with similar covariate values that are mapped to different latent variable values corresponding to different power spectra.  For these two realizations, (b) displays the true log power spectra (red lines), log periodogram ordinates (gray points), and estimated log power spectra using the proposed Bayesian sum of trees model (blue lines) and the AdaptSPEC-X model (green lines).}
    \label{fig:contrast}
\end{figure}


The first setting represents an AR(1) process in which the coefficient varies smoothly across one covariate and abruptly across another. The second setting  contains complex linear and nonlinear covariate effects and interactions \citep{friedman1991} adapted for time series data. The third setting represents an abruptly-changing process over two dimensions, similar to that of \cite{BertolacciMichael2019ACDS}.

\subsection{Results: Estimation Accuracy}
Hyperparameters are specified as $\sigma_{\alpha}^{2}=100$ for the prior variance of $\alpha_b$ in Equation \eqref{eq:splinemodel} and $\xi_{\tau}=2$ and $A_{\tau}=10$ for the prior on $\tau_b$ in Equation \eqref{posterior_tau1} and Equation \eqref{posterior_tau2}.
We used $M=5$ trees for estimation, and different numbers ($L$) and lengths ($T$) of time series are considered. 
The MCMC procedure is run for a total of 10,000 iterations with the first 5,000 discarded as burn-in.  
In order to assess convergence, trace plots for summary measures of the mean squared residuals, estimated log power spectrum, and tree structures
for all settings are available in the Web Appendix B.  These diagnostics appear to show convergence after approximately 5,000 iterations across all settings.   

\begin{figure}
  \centering
    \includegraphics[scale=0.4]{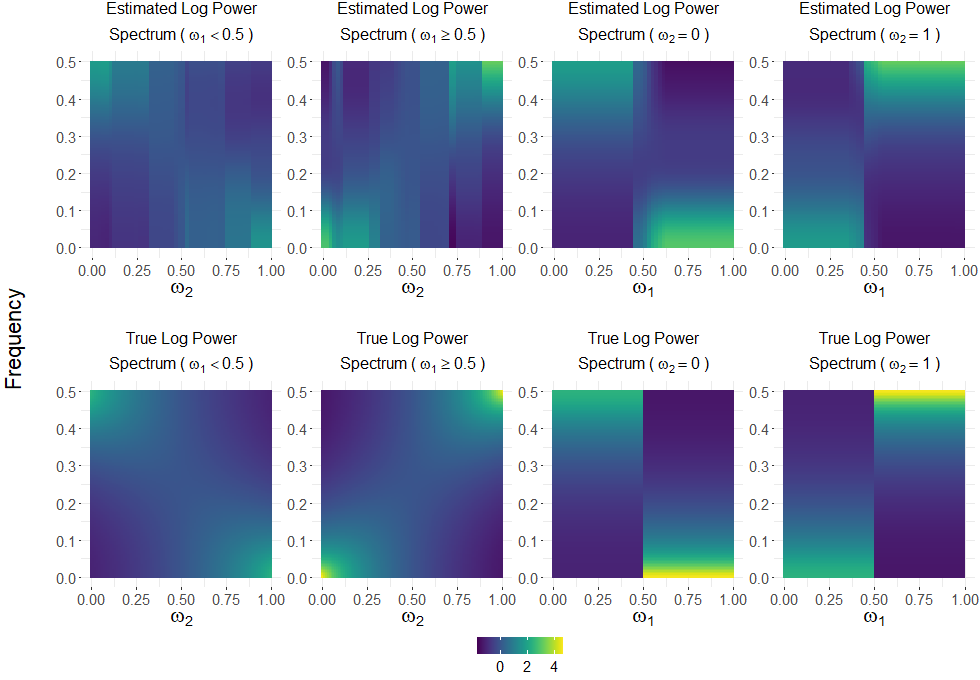}
   \caption{ Estimated and true covariate-dependent conditional log power spectrum for one run of the Abrupt+Smooth simulation setting.  The first two columns contain the estimated and true covariate-dependent power spectrum  conditional on $\omega_1 < 0.5$ and $\omega_1\geq 0.5$ respectively; The last two columns display the estimated and true covariate-dependent power spectrum  conditional on $\omega_2=0$ and $\omega_2=1$ respectively.}
  \label{fig:abrupt_slowly_estimation}
\end{figure}

Estimates of the covariate-dependent power spectrum for this run are presented in Figure \ref{fig:abrupt_slowly_estimation} to visually illustrate the ability of the proposed method to capture abrupt and smooth changes in power spectra. The smooth change in the conditional power spectrum over $\omega_2$ is captured by averaging over the posterior distribution of tree structures, which contains many splits across the range of possible values of $\omega_2$.  On the other hand, splits on $\omega_1$ are concentrated around the true abrupt change at $\omega_1=0.5$, and the posterior mean estimator of the conditional power spectrum over $\omega_1$ appropriately reflects the abrupt change in the conditional power spectrum.

To evaluate estimation accuracy, the mean and standard deviation of the mean squared error (MSE) is presented in Table \ref{table:MSE} for 100 replications of all settings. For comparison, AdaptSPEC-X \citep{BertolacciMichael2019ACDS} is also used to estimate the covariate-dependent power spectrum.  It is important to note that AdaptSPEC-X allows for modeling of time- and covariate-varying power spectra with time-varying means, which is a more general setting than what is considered in this work.  Accordingly, we implement a simplified version of the AdaptSPEC-X mixture model without the time-varying mean and power spectra components to enable more accurate comparisons.  We applied $C=50$ mixture components to ensure sufficient flexibility in estimating covariate effects on power spectra.

\begin{table}[ht!]
 \vspace*{-6pt}
 \centering
 \def\~{\hphantom{0}}
\caption{Mean and standard deviation of MSE over 100 replications for three simulation settings with different lengths ($T$) and number ($L$) of time series. Results are presented for the proposed Bayesian sum of trees model (top) with number of trees $M=5$ and the AdaptSPEC-X model (bottom) with mixture components $C=50$.}
 \label{table:MSE}
\begin{tabular*}{\columnwidth}{@{}c@{\extracolsep{\fill}}c@{\extracolsep{\fill}}c@{\extracolsep{\fill}}c@{\extracolsep{\fill}}c@{\extracolsep{\fill}}c@{\extracolsep{\fill}}c@{\extracolsep{\fill}}c@{\extracolsep{\fill}}c@{\extracolsep{\fill}}c@{}}
 \Hline
 & & \textbf{Abrupt+Smooth} & \textbf{AR-Friedman} & \textbf{Adj. AdaptSPEC-X}\\[1pt]
 
\cline{3-5} \\ [-6pt]
 {L} & {T} & \multicolumn{4}{c}{{Proposed Bayesian Sum of Trees Model}}\\ [1pt]

\hline
100 &  100 & 0.0406(0.0062) & 0.0386(0.0059) & 0.1475(0.0249)\\ 
 100 &  250 & 0.0217(0.0031)& 0.0226(0.0030) & 0.0762(0.0102)\\
 100 &  500 & 0.0138(0.0019)& 0.0153(0.0019)  & 0.0528(0.0063) \\
 200 &  100 & 0.0320(0.0052)& 0.0338(0.0042) & 0.1446(0.0202)\\ 
 200 &  250 & 0.0173(0.0030)& 0.0197(0.0020) & 0.0719(0.0070)\\ 
 200 &  500 & 0.0118(0.0023) & 0.0138(0.0015) & 0.0483(0.0050)\\ 
 500 &  100 & 0.0238(0.0030)& 0.0276(0.0022)&0.1348(0.0140)\\ 
 500 &  250 & 0.0124(0.0015)&0.0160(0.0013) & 0.0663(0.0077)\\ 
 500 &  500 & 0.0085(0.0013)  & 0.0119(0.0012) & 0.0452(0.0060)\\ 
\hline

 {L} & {T} &\multicolumn{3}{c|}{{AdaptSPEC-X Model}}\\
\hline
100 &  100 & 0.0576(0.0113) & 0.0529(0.0082)& 0.6334(0.1256)\\
 100 &  250 & 0.0437(0.0108) & 0.0406(0.0073) & 0.5595(0.1163)\\
 100 &  500 & 0.0412(0.0126) & 0.0377(0.0070) & 0.5394(0.1174)\\
 200 &  100 & 0.0510(0.0087) & 0.0474(0.0058) & 0.5182(0.0735)\\
 200 &  250 & 0.0398(0.0089) & 0.0372(0.0060) & 0.4453(0.0657)\\
 200 &  500 & 0.0378(0.0093) & 0.0365(0.0061) & 0.4240(0.0619)\\
 500 &  100 & 0.0466(0.0053) & 0.0398(0.0039) & 0.4631(0.0445)\\
 500 &  250 & 0.0379(0.0051) & 0.0323(0.0049) & 0.4012(0.0410)\\
500 &  500 & 0.0370(0.0050) & 0.0319(0.0047) & 0.3851(0.0369)\\

\hline
\end{tabular*}\vskip18pt
\end{table}

These results show that both methods see improved estimation accuracy as the number ($L$) and length ($T$) of time series increase.  However, the proposed method generally carries smaller MSEs than AdaptSPEC-X for comparable settings.  This can be partially attributed to the presence of abrupt changes across one or more covariates in all settings, which are better captured by the proposed tree-based approach.  To illustrate this point,
Figure \ref{fig:Edgepoints_Comparision} shows the estimated power spectra for two simulated time series with similar covariate values separated by an abrupt change from the Adjusted-AdaptSPEC-X setting.  As noted in \cite{BertolacciMichael2019ACDS}, the thin-plate Gaussian process prior on the mixture weights, while flexible, is both smooth and stationary.  Accordingly, the AdaptSPEC-X power spectrum estimates for these two time series are similar due to smoothing across similar covariate values induced by the thin-plate Gaussian process prior.  On the other hand, the proposed Bayesian sum-of-trees model is able to accurately distinguish the abrupt change in the power spectra. Moreover, averaging over the posterior distribution of trees enables the proposed method to recover smooth changes well (Figure \ref{fig:abrupt_slowly_estimation}).  
Taken together, this results in superior estimation accuracy across all settings seen in Table \ref{table:MSE}.

\subsection{Results: Computation Time}
\label{sec:sim_computationtime}
Simulations were carried out on a Windows 10 machine with an 8-core Intel i7 3.6 GHz processor and 64 GB RAM using \texttt{R} version 4.0.3 \citep{rcite}. The R code for implementing the proposed model is provided as a zip file in Supporting Information and is described in Web Appendix C.  Computationally-intensive aspects of the methodology are written in C++ using \texttt{RcppArmadillo} \citep{rcpparmadillo} for more efficient computation and reduced run times. Replications were run in parallel across six cores.  For the simulation settings considered herein, the mean run time for each tree update after burn-in ranges from 0.02 to 0.43 seconds, depending on the number ($L$) and length ($T$) of time series. The distribution of mean run times for each setting with different $L$ and $T$ and $M=5$ trees are visualized in the Web Appendix B.



While run times generally increase as the number and length of time series increase, length increases have a bigger impact on run times relative to increases in the number of time series. Increasing $T$ from 100 to 500 while holding $L$ constant increases mean run times by a multiple of approximately 3.5, while increasing $L$ from 100 to 500 while holding $T$ constant increases mean run times by a multiple of approximately 2.1. This is expected since the number of Fourier frequencies grows with the length of the series and increases both the number of terms being summed in the log Whittle likelihood and the dimension of the cosine basis used to approximate local log power spectra.  However, increasing the number of time series does not change the number of Fourier frequencies and only requires adding more terms in the log Whittle likelihood and posterior distribution for the spline coefficients, which is less computationally expensive.  See Equations \eqref{eq:Whittle}-\eqref{eq:Gibbsbeta} in Section \ref{sec:localest} and sampling scheme details in the Web Appendix A for more details.

Tree size also plays an important role in determining run times for tree updates.  Larger trees tend to have fewer time series belonging to each terminal node, which reduces computational burden and run times for evaluating modifications to a single terminal node.  This is why the Adjusted-AdaptSPEC-X setting, which requires larger trees to recover the complex covariate effects, has faster mean run times compared to other settings.  Additionally, tree size tends to increase as both $L$ and $T$ increase.  For illustration, the Web Appendix B includes a plot of the total number of bottom nodes for different combinations of $L$ and $T$ for the AR-Friedman simulation setting. This can also help explain why run times grow more slowly in $L$ compared to $T$.




\subsection{Results: Sparse Covariate Effects}
\label{sec:sparse}

In order to demonstrate the capability of the proposed method in providing efficient variable selection by adapting to sparse covariate effects for high-dimensional covariates, we consider a modification to the AR-Friedman simulation setting introduced previously. The original covariate vector containing 5 important covariates is now extended to include 95 additional noise covariates independently drawn from a standard normal distribution.  Variable selection efficiency can be investigated by assessing the estimated posterior probability for model inclusion of each covariate, which is the proportion of posterior draws where the covariate appears in at least one split rule for at least one tree. The estimated posterior probabilities for model inclusion for each of the 100 covariates using the uniform hyperprior and the sparsity-inducing Dirichlet prior for a single run of the proposed algorithm are shown in Web Appendix B.
Both forms of the prior correctly estimate the posterior probabilities of model inclusion to be 1 for all 5 important variables,
but the Dirichlet prior accurately achieves a sparser solution with the posterior probability of inclusion for the noise variables being much closer to 0.  Specifically, the mean posterior probability of inclusion for noise variables using the uniform prior is 0.11 with a standard deviation of 0.1861, and the mean posterior probability of inclusion for noise variables using the Dirichlet prior is 0.01 with a standard deviation of 0.1026.  For high-dimensional covariates in which only a few covariates are expected to be associated with the properties of the time series, the Dirichlet prior provides accurate selection of important covariates.

\section{Gait Maturation Analysis}
\label{sec:app}
We now present the analytical results of applying the proposed method to the motivating gait maturation study described in the introduction \citep{PhysioNet}. The current analysis considers the effect of age on gait variability to better understand gait maturation in young children in the presence of other factors that may influence gait, such as gender and gait speed.
The data contains stride interval time series from 50 healthy children with equal numbers of girls and boys between 3 and 14 years old.  
The time series  consist of $T=256$ stride times during normal walking after removing the first 60 seconds and last 5 seconds to avoid warm-up and ending effects (Figure \ref{fig:Timeseries}). More details of data processing can be found in \cite{HausdorffJ.M1999Mogd}.  The proposed Bayesian sum of trees model was used to estimate the covariate-dependent power spectrum of stride interval time series using 5 trees and 10,000 total iterations with the first 5,000 iterations discarded as burn-in. See Web Appendix B for convergence diagnostics of this application.





Partial dependence (PD) \citep{JeromeH.Friedman2001GFAA} is the most widely used method for evaluating covariate effects in machine learning models. However, there is an issue with multicollinearity in this dataset, as age and gait speed are significantly correlated $(r=0.653, p<0.0001)$, which can render PD unreliable due to extrapolation of the response at predictor values far outside the multivariate envelope of the data \citep{ApleyDanielW2020Vteo}. 
Therefore, we use accumulated local effects (ALE) \citep{ApleyDanielW2020Vteo} to characterize covariate effects. ALE presents the effect in a small interval of the interested feature, which can mitigate issues with multicollinearity by localizing the estimated effect within the envelope of the data.  
Let $\boldsymbol{\omega} = (\omega_j,\boldsymbol{\omega}_{\setminus j})$ where $\omega_j$ denotes the $j$th covariate and $\boldsymbol{\omega}_{\setminus j}$ denotes all covariates other than the $j$th covariate.  The ALE for $\omega_{j}=x$  on the power spectrum at frequency $\nu$ is defined as
\begin{equation}\label{eq:true_ALE}
\begin{split}
f_{j,\text{ALE}}(x,\nu) &= \int_{z_{0,j}}^{x} E_{\boldsymbol{\omega}_{\setminus j}|\omega_j}\left[\frac{\delta f(\boldsymbol{\omega},\nu)}{\delta \omega_{j}} \bigg| \omega_{j}=z_{j}\right]dz_{j}-\text{constant}
\end{split}
\end{equation}
where 
$\mathbb{Z}_j=\{z_{0,j},\ldots,z_{H,j}\}$ is a collection of $H+1$ partition points over the effective support of $\omega_{j}$. The constant is a value to vertically center the plot. Let $\hat{f}(z_{h,j},x_{\setminus j};\nu)$ be the estimated power spectrum for $\omega_{j}=z_{h,j}$, $h=1,\ldots,H$ and $\boldsymbol{\omega}_{\setminus j}=x_{\setminus j}$ on frequency $\nu$, the uncentered ALE can then be estimated by
\begin{equation}\label{eq:estimated_ALE}
    \hat{g}_{j,\text{ALE}}(x,\nu)=\sum_{h=1}^{h_j(x)}\frac{1}{n_{j}(h)}\sum_{\{i:x_{j}^{(i)}\in N_{j}(h)\} }\Big[\hat{f}(z_{h,j},x_{\setminus j}^{(i)};\nu)-\hat{f}(z_{h-1,j},x_{\setminus j}^{(i)};\nu) \Big]
\end{equation}
 where 
$h_{j}(x)$ is the index for the interval of the partition $\mathbb{Z}_j$ to which the value $x$ belongs, 
$n$ is the total number of observations and $n_{j}(h)$ is the number of observations in the $h$th segment of the partition for $\omega_{j}$ such that $\sum_{h=1}^{H}n_{j}(h)=n$. $N_{j}(h)=(z_{h-1,j},z_{h,j}]$ represents the $h$th interval of the partition for $\omega_{j}$. Then the estimated centered ALE is
\begin{equation}\label{eq:estimated_ALEcent}
    \hat{f}_{j,\text{ALE}}(x,\nu)=\hat{g}_{j,\text{ALE}}(x,\nu)-\frac{1}{n}\sum_{i=1}^{n}\hat{g}_{j,\text{ALE}}(x^{(i)},\nu).
\end{equation}

\begin{figure}
    \centering

    \subfigure[\label{fig:ALE_age_spectrum}]{\includegraphics[width=0.37\textwidth]{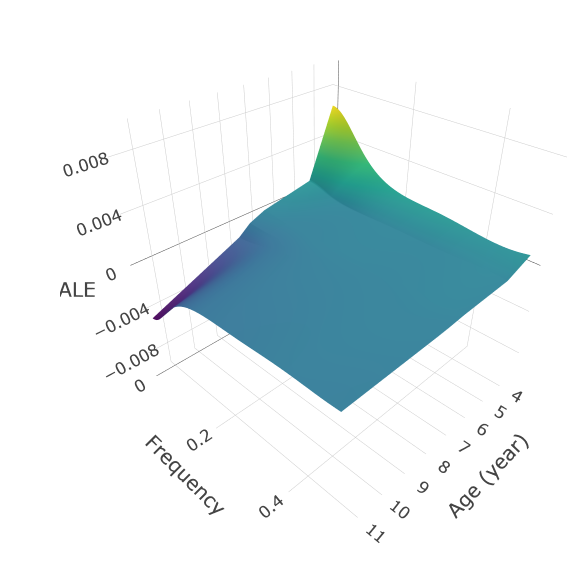}} 
    \subfigure[\label{fig:ALE_speed_spectrum}]{\includegraphics[width=0.37\textwidth]{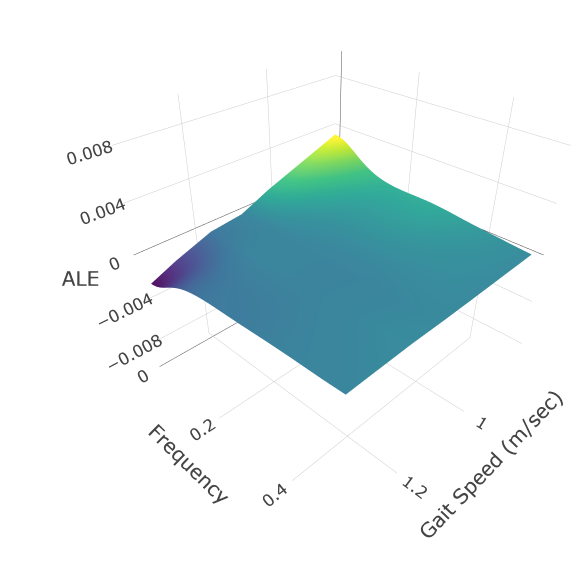}}\\
      \subfigure[\label{fig:ALE_age_ratio}]{\includegraphics[width=0.35\textwidth]{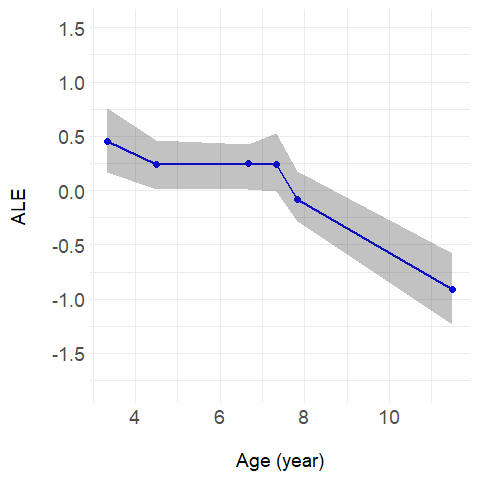}} 
    \subfigure[\label{fig:ALE_speed_ratio}]{\includegraphics[width=0.35\textwidth]{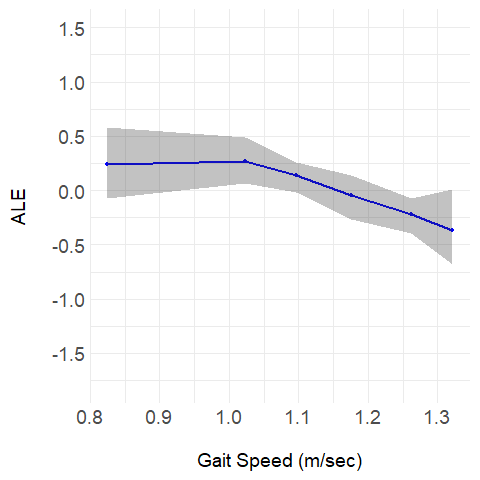}}

    \caption{Posterior mean of ALE for age (a) and gait speed (b) effects on the power spectrum and posterior mean of ALE for age (c) and gait speed (d) on LF/HF ratio (blue dotted line) with 95\% pointwise credible intervals (shaded gray region).}
    \label{fig:ALE}
\end{figure}

By partitioning covariates into $H=5$ intervals containing equal numbers of observations, Figures \ref{fig:ALE_age_spectrum} and \ref{fig:ALE_speed_spectrum} show the posterior mean of the ALE for age and gait speed on the power spectrum. Two findings can be concluded from these plots. First, power over all frequencies decreases as age increases. 
This indicates variability in stride times decreases with age, which is consistent with previous findings \citep{HausdorffJ.M1999Mogd}. Second, we can observe that power in low frequencies (LF) (0.05-0.25 stride$^{-1}$) decreases much more with age relative to higher frequencies (HF) (0.25-0.5 stride$^{-1}$).  This is also expected as low frequency power corresponds to fluctuations over relatively longer time scales and is indicative of less mature neuromuscular control \citep{HausdorffJ.M1999Mogd}.  To further illustrate this point, Figures \ref{fig:ALE_age_ratio} and \ref{fig:ALE_speed_ratio} present the posterior mean of the ALE of age and gait speed on the LF/HF ratio
\begin{equation*}
    \frac{\text{LF}}{\text{HF}}(\boldsymbol{\omega})=\frac{\int_{0.05}^{0.25}f(\boldsymbol{\omega},\nu)d\nu}{\int_{0.25}^{0.5}f(\boldsymbol{\omega},\nu)d\nu}
\end{equation*}
along with 95\% pointwise credible intervals. This can be computed by replacing the power spectrum $f(\boldsymbol{\omega},\nu)$ and estimated power spectrum $\hat{f}(z_{h,j},x_{\setminus j};\nu)$ in Equations \eqref{eq:true_ALE}-\eqref{eq:estimated_ALEcent} with the $\frac{\text{LF}}{\text{HF}}(\boldsymbol{\omega})$ and its corresponding estimates
\begin{equation*}
    {\frac{\widehat{\text{LF}}}{\text{HF}}}(\boldsymbol{\omega})=\frac{\sum_{\nu_k \in (0.05,0.25)}\hat{f}(\boldsymbol{\omega},\nu_k)}{\sum_{\nu_k \in (0.25,0.5)}\hat{f}(\boldsymbol{\omega},\nu_k)}
\end{equation*}
where $\frac{\widehat{\text{LF}}}{\text{HF}}(\boldsymbol{\omega})$ can be expressed as $\frac{\widehat{\text{LF}}}{\text{HF}}(z_{h,j},x_{\setminus j})$ when calculating the ALE for the $j$th covariate.  While $\text{LF/HF}(\omega)$ decreases significantly with both age and gait speed, we see relatively larger decreases beyond 7 years of age and for speeds above 1 m/sec. This is also consistent with previous findings \citep{HausdorffJ.M1999Mogd}, showing the LF/HF ratio does not show a significant decrease in children 3-7 years of age, but does show a significant decrease in children 7-14 years of age. Noting that gait speed is positively correlated with age, it is expected that speed exhibits similar association with the power spectrum.  Our analysis (not shown) did not indicate a significant gender effect on either the stride interval power spectrum or LF/HF ratio.

\section{Discussion}
\label{sec:end}

This paper describes a novel adaptive Bayesian covariate-dependent model for the power spectrum of multiple time series. By using a sum of trees model to characterize covariate effects, this model is flexible and can automatically recover complex nonlinear associations and interactions as well as provide efficient variable selection. 

This article is one of the first approaches to analyzing the power spectrum of multiple stationary time series with multiple covariates in a completely nonparametric manner, but it is not without limitations.  An important direction for future research is to extend the proposed Bayesian sum of trees model to accommodate additional data features commonly encountered in practice, such as time- and covariate-dependent time series \citep{BertolacciMichael2019ACDS}, extra spectral variability due to clustering effects \citep{Krafty2016}, and missingness in covariate vectors. Alternative partitioning frameworks, such Voronoi tesselations \citep{payne2020conditional} and binary space partitioning trees \citep{fan2019binary}, may also be considered for capturing covariate effects in an even more flexible and parsimonious manner.

\section*{Acknowledgements}

This work is supported by the National Institute of General Medical Sciences of the National Institutes of Health under Award Number R01GM140476.\vspace*{-8pt}


%
\bibliographystyle{biom} 
\bibliography{locstatbib.bib}

\begin{thebibliography}{}

\bibitem[\protect\citeauthoryear{Adak}{Adak}{1998}]{Adak1998}
Adak, S. (1998).
\newblock Time-dependent spectral analysis of nonstationary time series.
\newblock {\em Journal of the American Statistical Association} {\bf 93,} pp.
  1488\--1501.

\bibitem[\protect\citeauthoryear{Apley and Zhu}{Apley and
  Zhu}{2020}]{ApleyDanielW2020Vteo}
Apley, D.~W. and Zhu, J. (2020).
\newblock Visualizing the effects of predictor variables in black box
  supervised learning models.
\newblock {\em Journal of the Royal Statistical Society. Series B, Statistical
  Methodology} {\bf 82,} 1059--1086.

\bibitem[\protect\citeauthoryear{Bertolacci, Rosen, Cripps, and
  Cripps}{Bertolacci et~al.}{2019}]{BertolacciMichael2019ACDS}
Bertolacci, M., Rosen, O., Cripps, E., and Cripps, S. (2019).
\newblock {AdaptSPEC-X}: Covariate dependent spectral modeling of multiple
  nonstationary time series.
\newblock {\em ArXiv} .

\bibitem[\protect\citeauthoryear{Blattenberger and Fowles}{Blattenberger and
  Fowles}{2017}]{BlattenbergerGail2017TAFM}
Blattenberger, G. and Fowles, R. (2017).
\newblock Treed avalanche forecasting: Mitigating avalanche danger utilizing
  {B}ayesian additive regression trees.
\newblock {\em Journal of Forecasting} {\bf 36,} 165--180.

\bibitem[\protect\citeauthoryear{Breiman}{Breiman}{2001}]{BreimanLeo2001RF}
Breiman, L. (2001).
\newblock Random forests.
\newblock {\em Machine Learning} {\bf 45,} 5--32.

\bibitem[\protect\citeauthoryear{Brenière and Bril}{Brenière and
  Bril}{1988}]{article}
Brenière, Y. and Bril, B. (1988).
\newblock Why do children walk when falling down while adults fall down in
  walking?
\newblock {\em Comptes Rendus de l'Académie des Sciences. Série III, Sciences
  de la Vie} {\bf 307,} 617--22.

\bibitem[\protect\citeauthoryear{Brillinger}{Brillinger}{2002}]{brillinger2001}
Brillinger, D.~R. (2002).
\newblock {\em Time Series: Data Analysis and Theory}.
\newblock Philadelphia: SIAM.

\bibitem[\protect\citeauthoryear{Bruce, Hall, Buysse, and Krafty}{Bruce
  et~al.}{2018}]{BruceScottA2018CaBs}
Bruce, S.~A., Hall, M.~H., Buysse, D.~J., and Krafty, R.~T. (2018).
\newblock Conditional adaptive {B}ayesian spectral analysis of nonstationary
  biomedical time series.
\newblock {\em Biometrics} {\bf 74,} 260--269.

\bibitem[\protect\citeauthoryear{Chau and von Sachs}{Chau and von
  Sachs}{2016}]{chauandsachs2016}
Chau, J. and von Sachs, R. (2016).
\newblock {Functional mixed effects wavelet estimation for spectra of
  replicated time series}.
\newblock {\em Electronic Journal of Statistics} {\bf 10,} 2461 -- 2510.

\bibitem[\protect\citeauthoryear{Chipman, George, Gramacy, and
  McCulloch}{Chipman et~al.}{2013}]{Chipman2013}
Chipman, H., George, E.~I., Gramacy, R.~B., and McCulloch, R. (2013).
\newblock {B}ayesian treed response surface models.
\newblock {\em WIREs Data Mining and Knowledge Discovery} {\bf 3,} 298--305.

\bibitem[\protect\citeauthoryear{Chipman, George, and McCulloch}{Chipman
  et~al.}{2010}]{ChipmanHughA2010BBar}
Chipman, H.~A., George, E.~I., and McCulloch, R.~E. (2010).
\newblock {{BART}}: {B}ayesian additive regression trees.
\newblock {\em The Annals of Applied Statistics} {\bf 4,} 266--298.

\bibitem[\protect\citeauthoryear{Cram\'{e}r}{Cram\'{e}r}{1942}]{cramer42}
Cram\'{e}r, H. (1942).
\newblock On harmonic analysis in certain function spaces.
\newblock {\em Arkiv f\"{o}r Matematik, Astronomioch Fysik} {\bf 28B,} 1--7.

\bibitem[\protect\citeauthoryear{Diggle and Al~Wasel}{Diggle and
  Al~Wasel}{1997}]{diggle1997spectral}
Diggle, P.~J. and Al~Wasel, I. (1997).
\newblock Spectral analysis of replicated biomedical time series.
\newblock {\em Journal of the Royal Statistical Society: Series C (Applied
  Statistics)} {\bf 46,} 31--71.

\bibitem[\protect\citeauthoryear{Eddelbuettel and Sanderson}{Eddelbuettel and
  Sanderson}{2014}]{rcpparmadillo}
Eddelbuettel, D. and Sanderson, C. (2014).
\newblock {RcppArmadillo}: Accelerating {R} with high-performance {C++} linear
  algebra.
\newblock {\em Computational Statistics and Data Analysis} {\bf 71,}
  1054--1063.

\bibitem[\protect\citeauthoryear{Eubank}{Eubank}{1999}]{eubank1999}
Eubank, R. (1999).
\newblock {\em Nonparametric Regression and Spline Smoothing}.
\newblock Boca Raton: CRC Press.

\bibitem[\protect\citeauthoryear{Fan, Li, and Sisson}{Fan
  et~al.}{2019}]{fan2019binary}
Fan, X., Li, B., and Sisson, S. (2019).
\newblock Binary space partitioning forest.
\newblock In {\em The 22nd International Conference on Artificial Intelligence
  and Statistics}, pages 3022--3031. PMLR.

\bibitem[\protect\citeauthoryear{Fiecas and Ombao}{Fiecas and
  Ombao}{2017}]{FiecasMark2017MtEo}
Fiecas, M. and Ombao, H. (2017).
\newblock Modeling the evolution of dynamic brain processes during an
  associative learning experiment.
\newblock {\em Journal of the American Statistical Association} {\bf 111,}
  1440--1453.

\bibitem[\protect\citeauthoryear{Friedman}{Friedman}{1991}]{friedman1991}
Friedman, J.~H. (1991).
\newblock Multivariate adaptive regression splines.
\newblock {\em The Annals of Statistics} {\bf 19,} 1--67.

\bibitem[\protect\citeauthoryear{Friedman}{Friedman}{2001}]{JeromeH.Friedman2001GFAA}
Friedman, J.~H. (2001).
\newblock Greedy function approximation: A gradient boosting machine.
\newblock {\em The Annals of Statistics} {\bf 29,} 1189--1232.

\bibitem[\protect\citeauthoryear{Gelman}{Gelman}{2006}]{gelman2006}
Gelman, A. (2006).
\newblock Prior distributions for variance parameters in hierarchical models
  (comment on article by {B}rowne and {D}raper).
\newblock {\em {B}ayesian Anal.} {\bf 1,} 515--534.

\bibitem[\protect\citeauthoryear{Goldberger, Amaral, Glass, Hausdorff, Ivanov,
  Mark, Mietus, Moody, Peng, and Stanley}{Goldberger et~al.}{2000}]{PhysioNet}
Goldberger, A.~L., Amaral, L. A.~N., Glass, L., Hausdorff, J.~M., Ivanov,
  P.~C., Mark, R.~G., Mietus, J.~E., Moody, G.~B., Peng, C.-K., and Stanley,
  H.~E. (2000).
\newblock {PhysioBank, PhysioToolkit, and PhysioNet}: Components of a new
  research resource for complex physiologic signals.
\newblock {\em Circulation} {\bf 101,} 215--220.

\bibitem[\protect\citeauthoryear{Green}{Green}{1995}]{green95}
Green, P.~J. (1995).
\newblock Reversible jump markov chain monte carlo computation and {B}ayesian
  model determination.
\newblock {\em Biometrika} {\bf 82,} 711--732.

\bibitem[\protect\citeauthoryear{Hall, Vasko, Buysse, Ombao, Chen, Cashmere,
  Kupfer, and Thayer}{Hall et~al.}{2004}]{Halletal2004}
Hall, M., Vasko, R., Buysse, D., Ombao, H., Chen, Q., Cashmere, J.~D., Kupfer,
  D., and Thayer, J.~F. (2004).
\newblock Acute stress affects heart rate variability during sleep.
\newblock {\em Psychosomatic Medicine} {\bf 66,} 56\--62.

\bibitem[\protect\citeauthoryear{Hausdorff, Zemany, Peng, and
  Goldberger}{Hausdorff et~al.}{1999}]{HausdorffJ.M1999Mogd}
Hausdorff, J.~M., Zemany, L., Peng, C.-K., and Goldberger, A.~L. (1999).
\newblock Maturation of gait dynamics: Stride-to-stride variability and its
  temporal organization in children.
\newblock {\em Journal of Applied Physiology} {\bf 86,} 1040--1047.

\bibitem[\protect\citeauthoryear{Iannaccone and Coles}{Iannaccone and
  Coles}{2001}]{iannaccone2001semiparametric}
Iannaccone, R. and Coles, S. (2001).
\newblock Semiparametric models and inference for biomedical time series with
  extra-variation.
\newblock {\em Biostatistics} {\bf 2,} 261--276.

\bibitem[\protect\citeauthoryear{Klimesch}{Klimesch}{1999}]{Klimesch1999}
Klimesch, W. (1999).
\newblock {EEG} alpha and theta oscillations reflect cognitive and memory
  performance: {A} review and analysis.
\newblock {\em Brain Research Reviews} {\bf 29,} 169\--195.

\bibitem[\protect\citeauthoryear{Krafty}{Krafty}{2016}]{Krafty2016}
Krafty, R.~T. (2016).
\newblock Discriminant analysis of time series in the presence of within-group
  spectral variability.
\newblock {\em Journal of Time Series Analysis} {\bf 37,} 435--450.

\bibitem[\protect\citeauthoryear{Krafty, Hall, and Guo}{Krafty
  et~al.}{2011}]{krafty2011}
Krafty, R.~T., Hall, M., and Guo, W. (2011).
\newblock Functional mixed effects spectral analysis.
\newblock {\em Biometrika} {\bf 98,} 583--598.

\bibitem[\protect\citeauthoryear{Krafty, Rosen, Stoffer, Buysse, and
  Hall}{Krafty et~al.}{2017}]{krafty2017}
Krafty, R.~T., Rosen, O., Stoffer, D.~S., Buysse, D.~J., and Hall, M.~H.
  (2017).
\newblock Conditional spectral analysis of replicated multiple time series with
  application to nocturnal physiology.
\newblock {\em Journal of the American Statistical Association} {\bf 112,}
  1405--1216.

\bibitem[\protect\citeauthoryear{Li, Bruce, Wutzke, and Long}{Li
  et~al.}{2021}]{LiBruce2020MultiCABS}
Li, Z., Bruce, S.~A., Wutzke, C.~J., and Long, Y. (2021).
\newblock Conditional adaptive {B}ayesian spectral analysis of replicated
  multivariate time series.
\newblock {\em Statistics in Medicine} {\bf 40,} 1989--2005.

\bibitem[\protect\citeauthoryear{Li and Krafty}{Li and
  Krafty}{2019}]{li2019adaptive}
Li, Z. and Krafty, R.~T. (2019).
\newblock Adaptive {B}ayesian time-frequency analysis of multivariate time
  series.
\newblock {\em Journal of the American Statistical Association} {\bf 114,}
  453--465.

\bibitem[\protect\citeauthoryear{Linero}{Linero}{2018}]{LineroAntonioR2018BRTf}
Linero, A.~R. (2018).
\newblock {B}ayesian regression trees for high-dimensional prediction and
  variable selection.
\newblock {\em Journal of the American Statistical Association} {\bf 113,}
  626--636.

\bibitem[\protect\citeauthoryear{Payne, Guha, Ding, and Mallick}{Payne
  et~al.}{2020}]{payne2020conditional}
Payne, R.~D., Guha, N., Ding, Y., and Mallick, B.~K. (2020).
\newblock A conditional density estimation partition model using logistic
  gaussian processes.
\newblock {\em Biometrika} {\bf 107,} 173--190.

\bibitem[\protect\citeauthoryear{Preis, Klemms, and Müller}{Preis
  et~al.}{2008}]{PreisSabine2008Gabm}
Preis, S., Klemms, A., and Müller, K. (2008).
\newblock Gait analysis by measuring ground reaction forces in children:
  Changes to an adaptive gait pattern between the ages of one and five years.
\newblock {\em Developmental Medicine and Child Neurology} {\bf 39,} 228--233.

\bibitem[\protect\citeauthoryear{Qin, Guo, and Litt}{Qin
  et~al.}{2009}]{QinLi2009ATFM}
Qin, L., Guo, W., and Litt, B. (2009).
\newblock A time-frequency functional model for locally stationary time series
  data.
\newblock {\em Journal of Computational and Graphical Statistics} {\bf 18,}
  675--693.

\bibitem[\protect\citeauthoryear{{R Core Team}}{{R Core Team}}{2021}]{rcite}
{R Core Team} (2021).
\newblock {\em R: A Language and Environment for Statistical Computing}.
\newblock R Foundation for Statistical Computing, Vienna, Austria.

\bibitem[\protect\citeauthoryear{Rosen, Wood, and Stoffer}{Rosen
  et~al.}{2012}]{rosen2012}
Rosen, O., Wood, S., and Stoffer, D. (2012).
\newblock {AdaptSPEC}: Adaptive spectral estimation for nonstationary time
  series.
\newblock {\em Journal of the American Statistical Association} {\bf 107,}
  1575--1589.

\bibitem[\protect\citeauthoryear{Ro\v{c}kov\'a and Saha}{Ro\v{c}kov\'a and
  Saha}{2019}]{RockovaVeronika2018OTfB}
Ro\v{c}kov\'a, V. and Saha, E. (2019).
\newblock On theory for {BART}.
\newblock In {\em Proceedings of Machine Learning Research}, volume~89 of {\em
  Proceedings of Machine Learning Research}, pages 2839--2848. PMLR.

\bibitem[\protect\citeauthoryear{Schwarz and Krivobokova}{Schwarz and
  Krivobokova}{2016}]{SchwarzKatsiaryna2016Auff}
Schwarz, K. and Krivobokova, T. (2016).
\newblock A unified framework for spline estimators.
\newblock {\em Biometrika} {\bf 103,} 121--131.

\bibitem[\protect\citeauthoryear{Shumway-Cook and Williams}{Shumway-Cook and
  Williams}{1995}]{VanderLindenDarlW1996SAWM}
Shumway-Cook, A. and Williams, M.~W. (1995).
\newblock Motor control: Theory and practical applications.

\bibitem[\protect\citeauthoryear{Starling, Murray, Carvalho, Bukowski, Scott,
  et~al\mbox{.}}{Starling et~al.}{2020}]{starling2018bart}
Starling, J.~E., Murray, J.~S., Carvalho, C.~M., Bukowski, R.~K., Scott, J.~G.,
  et~al. (2020).
\newblock {{BART}} with targeted smoothing: {A}n analysis of patient-specific
  stillbirth risk.
\newblock {\em Annals of Applied Statistics} {\bf 14,} 28--50.

\bibitem[\protect\citeauthoryear{Stoffer, Han, Qin, and Guo}{Stoffer
  et~al.}{2010}]{stoffer2010smoothing}
Stoffer, D.~S., Han, S., Qin, L., and Guo, W. (2010).
\newblock Smoothing spline {ANOPOW}.
\newblock {\em Journal of Statistical Planning and Inference} {\bf 140,}
  3789--3796.

\bibitem[\protect\citeauthoryear{van~der Merwe}{van~der
  Merwe}{2018}]{vanderMerweSean2018TSAo}
van~der Merwe, S. (2018).
\newblock Time series analysis of the southern oscillation index using
  {B}ayesian additive regression trees.
\newblock {\em ArXiv} .

\bibitem[\protect\citeauthoryear{Waldmann}{Waldmann}{2016}]{WaldmannPatrik2016GpuB}
Waldmann, P. (2016).
\newblock Genome-wide prediction using {B}ayesian additive regression trees.
\newblock {\em Genetics Selection Evolution (Paris)} {\bf 48,} 42--.

\bibitem[\protect\citeauthoryear{Wand, Ormerod, Padoan, and Frühwirth}{Wand
  et~al.}{2011}]{wand2011}
Wand, M.~P., Ormerod, J.~T., Padoan, S.~A., and Frühwirth, R. (2011).
\newblock Mean field variational {B}ayes for elaborate distributions.
\newblock {\em {B}ayesian Anal.} {\bf 6,} 847--900.

\bibitem[\protect\citeauthoryear{Whittle}{Whittle}{1952}]{Whittle1952}
Whittle, P. (1952).
\newblock The simultaneous estimation of a time series harmonic components and
  covariance structure.
\newblock {\em Trabajos de Estadistica} {\bf 3,} 43\--57.

\bibitem[\protect\citeauthoryear{Wu, Tjelmeland, and West}{Wu
  et~al.}{2007}]{WuYuhong2007BCPS}
Wu, Y., Tjelmeland, H., and West, M. (2007).
\newblock {B}ayesian {CART}: Prior specification and posterior simulation.
\newblock {\em Journal of Computational and Graphical Statistics} {\bf 16,}
  44--66.

\end{thebibliography}








\newpage
\section*{Web Appendix A: Sampling Scheme Details}
\label{sec:A}
The sampling scheme for the proposed Bayesian sum of trees model for the covariate-dependent power spectrum is presented in this section. Suppose we have $M$ trees for the sum of trees model.  Let $(U_{j},\Phi_{j})$ and $(U^{*}_{j},\Phi^{*}_{j})$ be the current and proposed tree structure and terminal node parameter estimates respectively for the $j$th tree, and let $\mathbf{R}_{j}$ denote the residuals of the log periodogram ordinates from the fit corresponding to the sum of all trees except the $j$th tree for all time series.  More details about the types of proposals developed in this work are provided herein.  For ease of exposition, the subscript $j$ is dropped in what follows. 
We implement the reversible jump Markov chain Monte Carlo (MCMC) sampling scheme by using a Metropolis-Hastings algorithm in which the acceptance ratio $\alpha$ is formulated as
\begin{equation*}\label{eq:acceptance_ratio}
    \alpha=\text{min}\big\{1, A \big\},
\end{equation*}
where
\begin{equation*}
    A=\frac{p(U^{*},\Phi^{*}|\mathbf{R}) \times q(U,\Phi|U^{*},\Phi^{*})}{p(U,\Phi|\mathbf{R}) \times q(U^{*},\Phi^{*}|U,\Phi)}.
\end{equation*}
\\
More details on the individual components of the acceptance ratio introduced above  are provided below. 

\subsubsection*{i. Distribution of $p(U,\Phi|\mathbf{R})$ and $p(U^{*},\Phi^{*}|\mathbf{R})$}
The joint posterior distribution $p(U,\Phi|\mathbf{R})$ can be expressed as a product of the following terms
\begin{equation*}
\begin{split}
p(U,\Phi|\mathbf{R}) & =p(\mathbf{R}|U,\Phi)\times p(\Phi|U)\times p(U)\\
& =p(\mathbf{R}|U,\Phi)\times p(\boldsymbol{\beta},\tau^{2}|U)\times p(U)\\
& =\underbrace{p(\mathbf{R}|U,\Phi)}_{%
    \let\scriptstyle\textstyle
    \substack{\text{likelihood}}}\times \underbrace{p(\boldsymbol{\beta}|U,\tau^{2})\times p(\tau^{2}|U)\times p(U)}_{%
    \let\scriptstyle\textstyle
    \substack{\text{prior}}},
\end{split}
\end{equation*}
\\
where the prior of $\Phi$ is determined by the joint prior of $(\boldsymbol{\beta},\tau^2)$. Here, $\boldsymbol{\beta}$ represents the intercept and basis coefficients in Equation (1) of the manuscript. Suppose there are $n_{b}$ observations in the $b$th terminal node of $j$th tree for $b=1,\ldots,B$, the likelihood can be expressed as the product of individual Whittle likelihoods 
\begin{multline}\label{appdix:likelihood}
p(\mathbf{R}|U,\Phi) =\prod_{b=1}^{B}p(\mathbf{R}_{b1},\ldots,\mathbf{R}_{bn_{b}}|f_{jb})\approx \\
\prod_{b=1}^{B}\prod_{i=1}^{n_{b}}(2\pi)^{-n/2}\prod_{k=1}^{n}\exp\{\log f_{jb}(\nu_{k})+\exp\big(\mathbf{R}_{bi}(\nu_{k})\big)/f_{jb}(\nu_{k})\},
\end{multline}
where $n=\lfloor T/2 \rfloor-1$, $\nu_{k}=k/T$ for $k=1,\ldots,n$ are the Fourier frequencies, $T$ is the length of time series, and $\mathbf{R}_{bi}$ is the residual of the log periodogram ordinates for the $i$th time series belonging to the $b$th terminal node. 

The prior for $\boldsymbol{\beta}$ is a normal distribution such that $p(\boldsymbol{\beta}|U,\tau^{2}) \sim N(0,(\sigma_{\alpha}^{2},\tau^{2}\boldsymbol{D}_{S}))$ as described in Section 3.2 of the manuscript. The prior distribution for $\tau^2$,  $p(\tau^{2}|U)$, is a half-t distribution. We follow \cite{wand2011} and express the half-t distribution as a scale mixture of inverse gamma distributions with latent variable $a$ such that
\begin{equation*}
    p(\tau^{2}|a) \sim \text{IG}\left(\frac{\xi_{\tau}}{2}, \frac{\xi_{\tau}}{a}\right), \quad p(a)\sim \text{IG}\left(\frac{1}{2}, \frac{1}{A_{\tau}^{2}} \right).
\end{equation*}
The prior for the tree structure $p(U)$ is 
\begin{equation*}\label{appdix:tree_structure}
\begin{split}
p(U) &=\prod_{\eta\in \text{terminals}}[1-p_{\text{split}}(\eta)]\prod_{\eta\in \text{internals}}p_{\text{split}}(\eta)\prod_{\eta\in \text{internals}}p_{\text{rule}}(\eta),
\end{split}
\end{equation*}
where $\eta$ is the node of the current tree. The $p_{\text{split}}(\eta)=\gamma(1+d)^{-\theta}$ is the probability of node $\eta$ to be split into two child nodes with $\gamma\in(0,1), \ \theta\in[0,\infty)$ and $d=0,1,\ldots$ which is the depth of the given node $\eta$. The $p_{\text{rule}}(\eta)=\frac{1}{n_{\text{adj}}(\eta)}\times\frac{1}{n_{\text{cutpoint}}(\eta)}$ is the probability of the available variables and the cutpoints to be chosen for node $\eta$, where  $n_{\text{adj}}(\eta)$ denotes the number of predictors available for the node and $n_{\text{cutpoint}}(\eta)$ is the number of available cutpoints for the selected variable. The posterior distribution $p(U^{*},\Phi^{*}|\mathbf{R})$ is similar with $p(U,\Phi|\mathbf{R})$ by plugging in $U^{*}$ and $\Phi^{*}$ instead.

\subsubsection*{ii. Distribution of $q(U^{*},\Phi^{*}|U,\Phi)$ and $q(U,\Phi|U^{*},\Phi^{*})$}
The proposed density $q(U^{*},\Phi^{*}|U,\Phi))$ is defined as
\begin{equation*} 
\begin{split}
q(U^{*},\Phi^{*}|U,\Phi) & = q(\Phi^{*}|U^{*},U,\Phi)\times q(U^{*}|U,\Phi) \\
& = q(\boldsymbol{\beta}^{*},\tau^{2*}|U^{*},U,\Phi)\times q(U^{*}|U)\\
& = q(\tau^{2*}|U^{*},U,\Phi)\times q(\boldsymbol{\beta}^{*}|\tau^{2*},U^{*},U,\Phi)\times q(U^{*}|U),
\end{split}
\end{equation*}
and similarly, the density $q(U,\Phi|U^{*},\Phi^{*})$ is
\begin{equation*}
\begin{split}
q(U,\Phi|U^{*},\Phi^{*}) & = q(\Phi|U^{*},\Phi^{*},U)\times q(U|U^{*},\Phi^{*}) \\
& = q(\boldsymbol{\beta},\tau^{2}|U^{*},\Phi^{*},U)\times q(U|U^{*})\\
& = q(\tau^{2}|U^{*},\Phi^{*},U)\times q(\boldsymbol{\beta}| \tau^{2},U^{*},\Phi^{*},U)\times q(U|U^{*}).
\end{split}
\end{equation*}
\\
From part \textbf{i} and part \textbf{ii}, $A$ can be written as
\begin{equation}\label{accept_ratio}
    \underbrace{\frac{p(\mathbf{R}|U^{*},\Phi^{*})}{p(\mathbf{R}|U,\Phi)}}_{%
    \let\scriptstyle\textstyle
    \substack{\text{likelihood ratio}}}
    \times
    \underbrace{    \frac{p(\boldsymbol{\beta}^{*}|U^{*},\tau^{2*})p(\tau^{2*}|U^{*})}{p(\boldsymbol{\beta}|U,\tau^{2})p(\tau^{2}|U)}}_{%
    \let\scriptstyle\textstyle
    \substack{\text{prior ratio}}}
    \times
    \underbrace{\frac{p(U^{*})}{p(U)}}_{%
    \let\scriptstyle\textstyle
    \substack{\text{tree structure}\\\text{ ratio}}}
    \times
    \underbrace{   \frac{q(\boldsymbol{\beta},\tau^{2}|U^{*},\Phi^{*},U)}{q(\boldsymbol{\beta}^{*},\tau^{2*}|U,\Phi,U^{*})}}_{%
    \let\scriptstyle\textstyle
    \substack{\text{proposed probability}\\ \text{ratio}}}
    \times
    \underbrace{   \frac{q(U|U^{*})}{q(U^{*}|U)}}_{%
    \let\scriptstyle\textstyle
    \substack{\text{transition ratio}}}.
\end{equation}

Proposed modifications to the tree structures can take the form of one of three possible moves: BIRTH, DEATH and CHANGE. The BIRTH move grows the tree by splitting a terminal node into two child nodes, the DEATH move prunes the tree by dropping two terminal child nodes belonging to the same internal node, and the CHANGE move modifies the variable and cut point associated with an internal node with two terminal child nodes.  Noticing that the prior ratio is the same for all BIRTH, DEATH, and CHANGE moves, the other ratios will be described individually for each of the three types of moves. 

\subsection*{BIRTH}
For the BIRTH move, the $b$th terminal node of the $j$th tree is selected to be split into two new child nodes. The proposed tree differs from the original tree only through the change from this terminal node to new child nodes. Therefore, the likelihood ratio becomes
\begin{equation*}
    \frac{p(\mathbf{R}|U^{*},\Phi^{*})}{p(\mathbf{R}|U,\Phi)}=\frac{p(\mathbf{R}_{b_l 1},\ldots,\mathbf{R}_{b_l n_{l}}|f_{jb_{l}})p(\mathbf{R}_{b_r 1},\ldots,\mathbf{R}_{b_r n_{r}}|f_{jb_{r}})}{p(\mathbf{R}_{b1},\ldots,\mathbf{R}_{bn_{b}}|f_{jb})},
\end{equation*}
where $b_l$ denotes left child node, $b_r$ is for the right child node, $n_l$ is the number of series corresponding to the proposed left child node, $n_r$ is the number of series corresponding to the proposed right child node, $\mathbf{R}_{b_l i}$ are residuals of the log periodogram ordinates for the $i$th series corresponding to the proposed left child node, and $\mathbf{R}_{b_r i}$ defined similarly for the proposed right child node. The likelihood within the terminal node is as shown in Equation \eqref{appdix:likelihood}. The tree structure ratio is expressed as
\begin{equation*}
\begin{split}
\frac{p(U^{*})}{p(U)} &=\frac{(1-p_{\text{split}}(b_{l}))(1-p_{\text{split}}(b_{r}))p_{\text{split}}(b)p_{\text{rule}}(b)}{1-p_{\text{split}}(b)}.
\end{split}
\end{equation*}
For the transition probability $q(U|U^{*})$ and $q(U^{*}|U)$, they can be expressed as
\begin{equation*} 
\begin{split}
q(U^{*}|U) & = p(\text{GROW})\times p(\text{selecting the $b$th terminal node to grow from})\\
& \times p(\text{selecting the $q$th predictor to split on})\\
& \times p(\text{selecting the $w$th value to split on})\\
 & = p(\text{GROW})\frac{1}{B}\frac{1}{n_{\text{adj}}(b)}\times\frac{1}{n_{\text{cutpoint}}(b)},
\end{split}
\end{equation*}
and
\begin{equation*} 
\begin{split}
q(U|U^{*}) & = p(\text{PRUNE})\times p(\text{selecting node $\eta$ to prune from}) \\
 & = p(\text{PRUNE})\frac{1}{n_{\text{internal}^{*}}},
\end{split}
\end{equation*}
where $p(\text{GROW})=0.25$ and $p(\text{PRUNE})=0.25$ are the probability of BIRTH and DEATH move to be selected, and $n_{\text{internal}^{*}}$ is the total number of internal nodes that has two terminal child nodes. Thus, we can derive the transition ratio to be
\begin{equation*}
    \frac{q(U|U^{*})}{p(U^{*}|U)}=\frac{p(\text{PRUNE})Bn_{\text{adj}}(b)n_{\text{cutpoint}}(b)}{q(\text{GROW})n_{\text{internal}^{*}}}.
\end{equation*}

Then, we follow \cite{rosen2012} to draw the proposed terminal parameters $\boldsymbol{\beta}^{*}$, $\tau^{2*}$ and latent variable $a^{*}$. For $\tau^{2}$(a), let $\tau^{2}_{b}(a_{b})$ denote the current $\tau^{2}(a)$ for the $b$th terminal node and a uniform distribution is used to generate new parameters $\tau^{2*}_{b_{l}}(a^{*}_{b_{l}})$ and $\tau^{2*}_{b_{r}}(a^{*}_{b_{r}})$ for the left and right child. Specifically,
\begin{equation*}
    \tau^{2*}_{b_{l}}=\tau^{2}_{b}\times\frac{u_{\tau}}{1-u_{\tau}}, \quad    \tau^{2*}_{b_{r}}=\tau^{2}_{b}\times\frac{1-u_{\tau}}{u_{\tau}}
\end{equation*}
\begin{equation*}
   a^{*}_{b_{l}}=a_{b}\times\frac{u_a}{1-u_a}, \quad    a^{*}_{b_{r}}=a_{b}\times\frac{1-u_a}{u_a}
\end{equation*}
where $u_{\tau}, u_a \sim U[0,1]$. For $\boldsymbol{\beta}$, an approximated normal distribution is proposed to generate the new parameters $\boldsymbol{\beta}^{*}_{b_{l}}$ and $\boldsymbol{\beta}^{*}_{b_{r}}$. Specifically, $(\boldsymbol{\beta}^{*}_{b^{*}} | \tau_{b^{*}}^{2*}, U^{*},U, \Phi) \sim N(\boldsymbol{\beta}^{\text{max}}_{b},\boldsymbol{\Sigma}^{\text{max}}_{b})$, where
\begin{equation}
    \boldsymbol{\beta}^{\text{max}}_{b}=\text{arg} \text{max}_{\boldsymbol{\beta}^{*}_{b^{*}}} \ p(\boldsymbol{\beta}^{*}_{b^{*}}|\mathbf{R}_{b},\tau^{2*}_{b^{*}},U^{*}),
\end{equation}
and
\begin{equation}
    \boldsymbol{\Sigma}^{\text{max}}_{b}=\biggm\{-(\partial^{2}\log p(\boldsymbol{\beta}^{*}_{b^{*}}|\mathbf{R}_{b},\tau^{2*}_{b^{*}},U^{*}))/(\partial \boldsymbol{\beta}^{*}_{b^{*}}\partial\boldsymbol{\beta}^{*'}_{b^{*}})|_{\boldsymbol{\beta}^{*}_{b^{*}}=\boldsymbol{\beta}^{\text{max}}_{b^{*}}}\biggm\}^{-1},
\end{equation}
where $p(\boldsymbol{\beta}^{*}_{b^{*}}|\mathbf{R}_{b},\tau^{2*}_{b^{*}},U^{*})$ is presented in Section 3.2, and $b^{*}$ represents the left or right child of node $b$. We then have the proposed probability ratio
\begin{equation*}
\begin{split}
\frac{q(\boldsymbol{\beta},\tau^{2}|U^{*},\Phi^{*},U)}{q(\boldsymbol{\beta}^{*},\tau^{2*}|U,\Phi,U^{*})} & = \frac{q(\boldsymbol{\beta}|\tau^{2},U^{*},\Phi^{*},U)q(\tau^{2}|U^{*},\Phi^{*},U)q(a|U^{*},\Phi^{*},U)}{q(\boldsymbol{\beta}^{*}|\tau^{2*}|U,\Phi,U^{*})q(\tau^{2*}|U^{*},\Phi^{*},U)q(a^*|U^{*},\Phi^{*},U)} \\
 & = \frac{q(\boldsymbol{\beta}_{b})}{q(\boldsymbol{\beta}^{*}_{b_{l}})q(\boldsymbol{\beta}^{*}_{b_{r}})p(u)}\times\biggm|\frac{\partial(\tau^{2*}_{b_{l}},\tau^{2*}_{b_{r}})}{\partial(\tau^{2}_{b},u_{\tau})}\biggm|\times\biggm|\frac{\partial(a^{*}_{b_{l}},a^{*}_{b_{r}})}{\partial(a_{b},u_a)}\biggm|,
 \end{split}
\end{equation*}
\\
where $p(u_{\tau}),p(u_a)=1$, $q(\boldsymbol{\beta}_b)$, $q(\boldsymbol{\beta}^{*}_{b_{l}})$ and $q(\boldsymbol{\beta}^{*}_{b_{r}})$ are the densities of the approximately normal distribution $N(\boldsymbol{\beta}^{\text{max}}_{b},\boldsymbol{\Sigma}^{\text{max}}_{b})$ with the corresponding current and proposed $\tau^{2}$ values. The Jacobian of $\tau^{2}$ and $a$ are
\begin{equation*}
    \biggm|\frac{\partial(\tau^{2*}_{b_{l}},\tau^{2*}_{b_{r}})}{\partial(\tau^{2}_{b},u_{\tau})}\biggm|=\frac{2\tau_{b}^{2}}{u_{\tau}(1-u_{\tau})}, \quad \biggm|\frac{\partial(a^{*}_{b_{l}},a^{*}_{b_{r}})}{\partial(a_{b},u_a)}\biggm|=\frac{2a_{b}}{u_{a}(1-u_{a})}.
\end{equation*}


\subsection*{DEATH} 
The DEATH move is the inverse of the BIRTH move. Suppose $b$ is the selected internal to be pruned by deleting the two child nodes $b_{l}$ and $b_{r}$, the likelihood ratio for the proposed tree and the current tree is hence expressed as
\begin{equation*}
    \frac{p(\mathbf{R}|U^{*},\Phi^{*})}{p(\mathbf{R}|U,\Phi)}=\frac{p(\mathbf{R}_{b1},\ldots,\mathbf{R}_{bn_{b}})|f_{jb}}{p(\mathbf{R}_{b_l,1},\ldots,\mathbf{R}_{b_l,n^{l}}|f_{jb_{l}})p(\mathbf{R}_{b_r,1},\ldots,\mathbf{R}_{b_r,n^{r}}|f_{jb_{r}})},
\end{equation*}
which is the change from the two child nodes to the internal $b$ node. The ratio of tree structure is presented as
\begin{equation*}
\begin{split}
\frac{p(U^{*})}{p(U)} &=\frac{1-p_{\text{split}}(b)}{(1-p_{\text{split}}(b_{l}))(1-p_{\text{split}}(b_{r}))p_{\text{split}}(b)p_{\text{rule}}(b)},
\end{split}
\end{equation*}
and the ratio of transition 
\begin{equation*}
    \frac{p(U|U^{*})}{p(U^{*}|U)}=\frac{q(\text{GROW})n_{\text{internal}^{*}}}{q(\text{PRUNE})(B-1)n_{\text{adj}}(b)n_{\text{cutpoint}}(b)},
\end{equation*}
where $B-1$ is the number of terminal nodes for the proposed prune tree. For the proposed probability ratio, we draw terminal node parameters  $\tau^{2*}_{b}$ and $a_{b}^{*}$ by taking the inverse of the corresponding BIRTH move
\begin{equation*}
    \tau^{2*}_{b}=\sqrt{\tau^{2}_{b_{l}}\tau^{2}_{b_{r}}}, \quad     a^{*}_{b}=\sqrt{a_{b_{l}}a_{b_{r}}}.
\end{equation*}
The vector $\boldsymbol{\beta}^{*}_{b}$ is drawn from the approximately normal distribution $N(\boldsymbol{\beta}^{\text{max}}_{b},\boldsymbol{\Sigma}^{\text{max}}_{b})$ given the proposed $\tau^{2*}$ value. Hence, the proposed probability ratio is
\begin{equation*}
\frac{q(\boldsymbol{\beta},\tau^{2}|U^{*},\Phi^{*},U)}{q(\boldsymbol{\beta}^{*},\tau^{2*}|U,\Phi,U^{*})} = \frac{q(\boldsymbol{\beta}_{b_{l}})q(\boldsymbol{\beta}_{b_{r}})p(u)}{q(\boldsymbol{\beta}_{b}^{*})}
\times
\biggm|\frac{\partial(\tau^{2*}_{b},u_{\tau})}{\partial(\tau^{2}_{b_{l}},\tau^{2}_{b_{r}})}\biggm|
\times
\biggm|\frac{\partial(a_{b}^{*},u_a)}{\partial(a_{b_{l}},a_{b_{r}})}\biggm|,
\end{equation*}
where the Jacobian of $\tau^{2}$ and $a$ are
\begin{equation*}
    \biggm|\frac{\partial(\tau^{2*}_{b},u_{\tau})}{\partial(\tau^{2}_{b_{l}},\tau^{2}_{b_{r}})}\biggm|=\frac{u_{\tau}(1-u_{\tau})}{2\tau_{b}^{2*}}=2(\tau_{b_l}+\tau_{b_r})^{2}, 
\end{equation*}
\begin{equation*}
    \biggm|\frac{\partial(a^{*}_{b},u_{a})}{\partial(a_{b_{l}},a_{b_{r}})}\biggm|=\frac{u_{a}(1-u_{a})}{2a_{b}^{*}}=2(\sqrt{a_{b_l}}+\sqrt{a_{b_r}})^{2}.
\end{equation*}

\subsection*{CHANGE} 
A CHANGE move is to change two terminal nodes to a pair of new child nodes by changing the split rule of their parent node. The observations within each terminal node of the proposed tree can be different from the current tree. Thus, the likelihood ratio is 
\begin{equation*}
    \frac{p(\mathbf{R}|U^{*},\Phi^{*})}{p(\mathbf{R}|U,\Phi)}=\frac{p(\mathbf{R}_{l,1},\ldots,\mathbf{R}_{l,n^{l^{*}}}|f_{jb_{l^{*}}})p(\mathbf{R}_{r,1},\ldots,\mathbf{R}_{r,n^{r^{*}}}|f_{jb_{r^{*}}})}{p(\mathbf{R}_{l,1},\ldots,\mathbf{R}_{l,n^{l}}|f_{jb_{l}})p(\mathbf{R}_{r,1},\ldots,\mathbf{R}_{r,n^{r}}|f_{jb_{r}})},
\end{equation*}
where $n^{l^{*}}$ and $n^{r^{*}}$ are the number of observations in the new left and the new right child nodes. The tree structure ratio for the CHANGE move is
\begin{equation*}
\begin{split}
\frac{p(U^{*})}{p(U)} &=\frac{(1-p_{\text{split}}(b_{l^{*}}))(1-p_{\text{split}}(b_{r^{*}}))p_{\text{split}}(b^{*})p_{\text{rule}}(b^{*})}{(1-p_{\text{split}}(b_{l}))(1-p_{\text{split}}(b_{r}))p_{\text{split}}(b)p_{\text{rule}}(b)},
\end{split}
\end{equation*}
since the depth of the children nodes does not change, the split probability for each node stays the same. The only change is the number of the available cutpoints of the new variable, which can be different from the current variable. Therefore, the tree structure ratio is
\begin{equation*}
    \frac{p(U^{*})}{p(U)}=\frac{n_{\text{cutpoint}}(b)}{n_{\text{cutpoint}}(b^{*})},
\end{equation*}
where $n_{\text{cutpoint}}(b^{*})$ is the number of available cutpoints for the proposed variable. The transition probability from current tree to the proposed tree is
\begin{equation*}
    \begin{split}
        p(U^{*}|U) & =p(\text{CHANGE})\times p(\text{selecting node $b$ to change})\\
        & \times p(\text{selecting the new predictor to split on})\\
        & \times p(\text{selecting the new cutpoint to split on}).
    \end{split}
\end{equation*}
The calculation of the $p(U|U^{*})$ is similar with $p(U^{*}|U)$ except that the number of available cutpoints can be different. So the transition ratio is
\begin{equation*}
    \frac{p(U|U^{*})}{p(U^{*}|U)}=\frac{n_{\text{cutpoint}}(b^{*})}{n_{\text{cutpoint}}(b)}.
\end{equation*}
From the results above, the transition ratio and the tree structure ratio are cancelled in the representation of Equation \eqref{accept_ratio}. So the acceptance ratio for the CHANGE step is only related to the likelihood, prior probability and the proposed probability of terminal node parameters. Two-step Gibbs sampling is used to draw new terminal node parameters. The $\boldsymbol{\beta}^{*}_{b_{*}}$ is drawn from the normal approximation $N(\boldsymbol{\beta}^{\text{max}}_{b},\boldsymbol{\Sigma}^{\text{max}}_{b})$ given the current $\tau_{b_{*}}^{2}$, and $a_{b^{*}}$ and $\tau_{b^{*}}^{2}$ are draw from their full conditional distributions. The MCMC algorithm  draws $a_{b^{*}}$ first and then updates $\tau_{b^{*}}^{2}$.

\section*{Web Appendix B:  Additional Results}
\label{sec:B}
Additional simulation results and the gait maturation analysis are shown in the following section. We set $M=5$, $L=100$, and $T=250$ when modeling the simulated data by using 10,000 iterations with the first 5,000 iterations discarded as burn-in. Also, we display the distribution of mean run times for the three simulations and the number of bottom nodes for different number of time series ($L$) and length of time series ($T$) of AR-Friedman simulation to illustrate the effect of $L$ and $T$ on the computation time.

\subsection*{Estimation and Convergence Diagnostics of Simulations}
\begin{figure}
  \centering
    \includegraphics[scale=0.65]{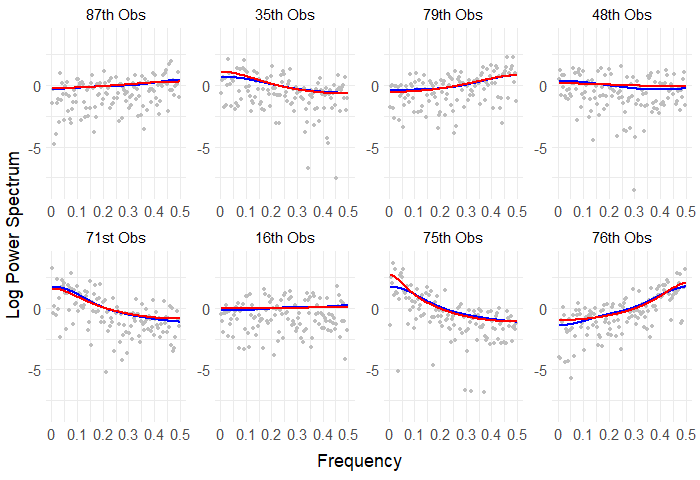}
   \caption{The estimated (blue line) and the true (red line) log power spectrum of the eight randomly selected time series for the AR-Friedman simulation. The corresponding log periodogram ordinates are shown with gray dots. }
  \label{fig:Est_friedman}
\end{figure}

\begin{figure}
  \centering
    \includegraphics[scale=0.65]{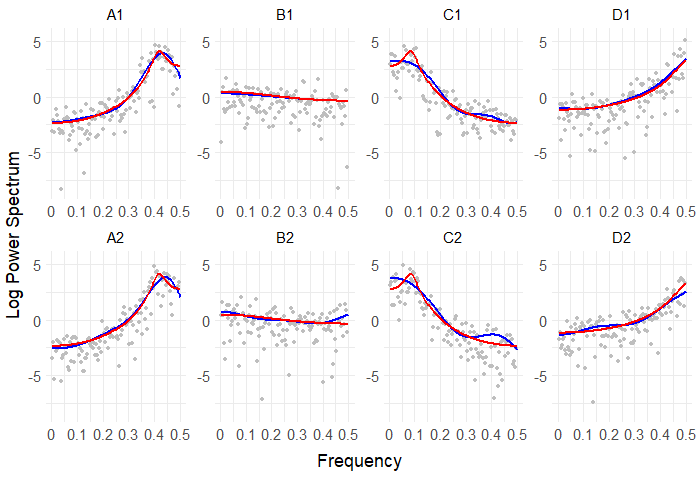}
   \caption{The estimated (blue line) and the true (red line) log power spectrum of the eight time series denoted in Figure \ref{fig:AdaptX_scatter}. The corresponding log periodogram ordinates are shown with gray dots. }
  \label{fig:Est_adaptX}
\end{figure}

\begin{figure}
  \centering
    \subfigure[]{\includegraphics[width=0.45\textwidth]{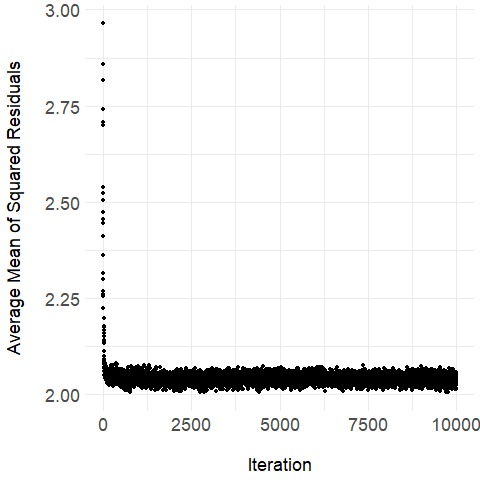}}
    \subfigure[]{\includegraphics[width=0.45\textwidth]{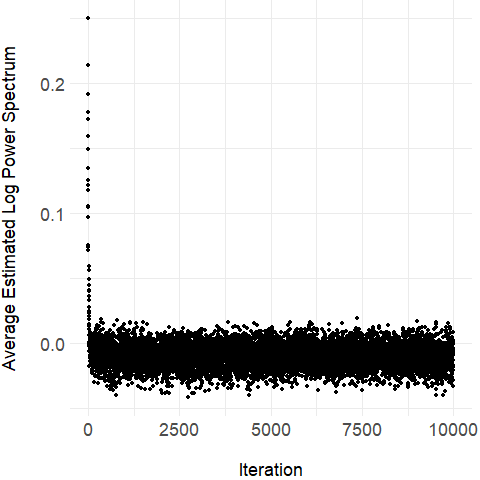}}
    \subfigure[]{\includegraphics[width=0.45\textwidth]{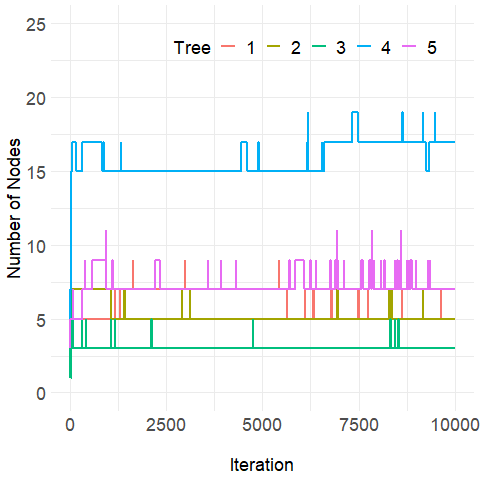}}
    \subfigure[]{\includegraphics[width=0.45\textwidth]{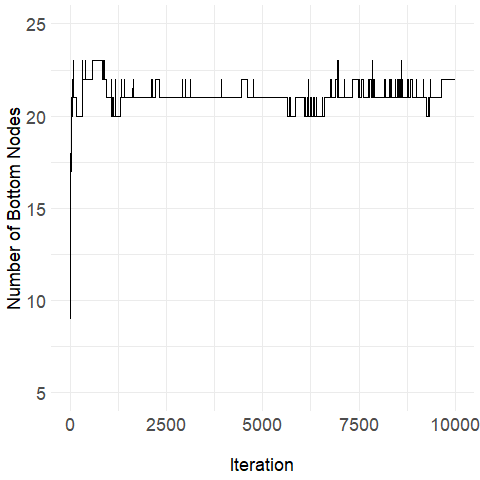}}
  \caption{
    Convergence diagnostic plots for the Abrupt+Smooth simulation for one replication: (a) average mean squared residuals across frequencies and all time series; (b) average estimated log power spectrum across frequencies and all time series; (c) total number of nodes for each of the five trees; (d) total number of bottom nodes across all five trees.}
  \label{fig:abrupt_diagnostic}
\end{figure}

\begin{figure}
  \centering
    \subfigure[Average mean of squared residuals]{\includegraphics[width=0.45\textwidth]{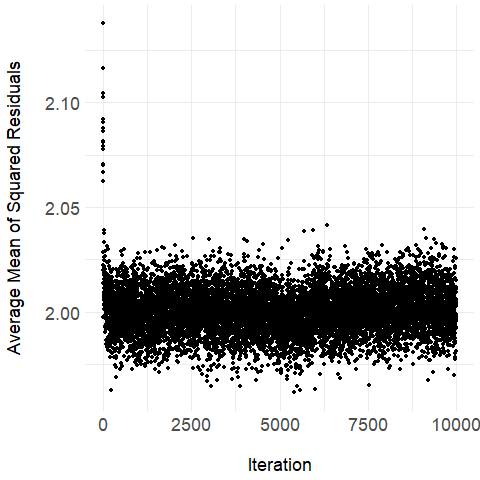}}
    \subfigure[Average of the estimated log spectrum]{\includegraphics[width=0.45\textwidth]{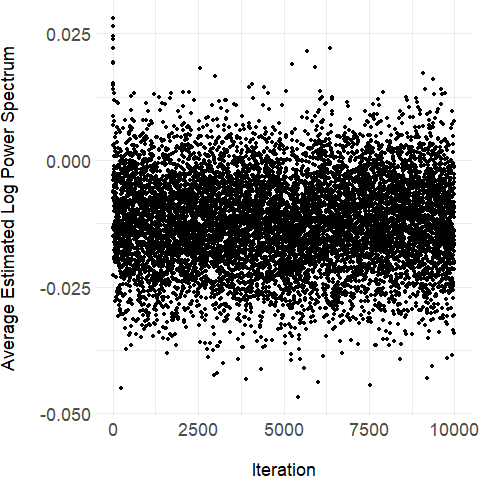}}
    \subfigure[Number of nodes of each tree]{\includegraphics[width=0.45\textwidth]{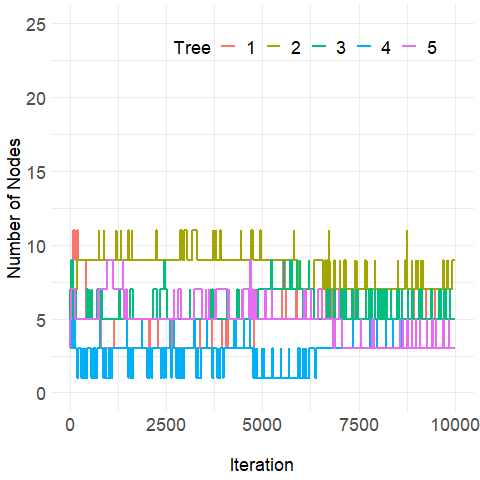}}
    \subfigure[Number of bottom nodes across all trees]{\includegraphics[width=0.45\textwidth]{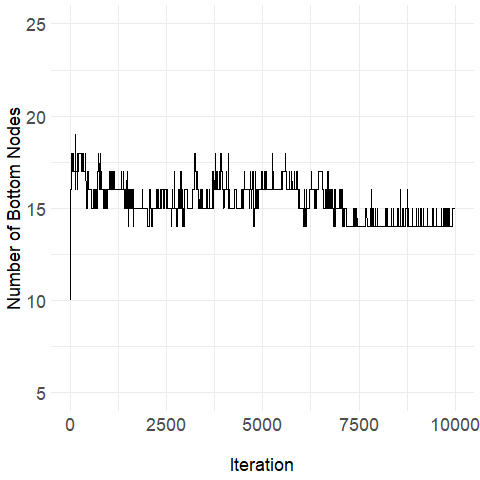}}
   \caption{
    Convergence diagnostics for the AR-Friedman simulation for one replication. Plot (a) contains trace plots of the average mean squared residuals across all time series; Plot (b) shows the average estimated log power spectrum across frequencies and all time series for each iteration; Plot (c) is the trace plots of the total number of nodes for each of the five trees separately (c); Plot (d) is the total number of bottom nodes across all five trees.}
  \label{fig:friedman_diagnostic}
\end{figure}

\begin{figure}
  \centering
    \subfigure[Average mean of squared residuals]{\includegraphics[width=0.45\textwidth]{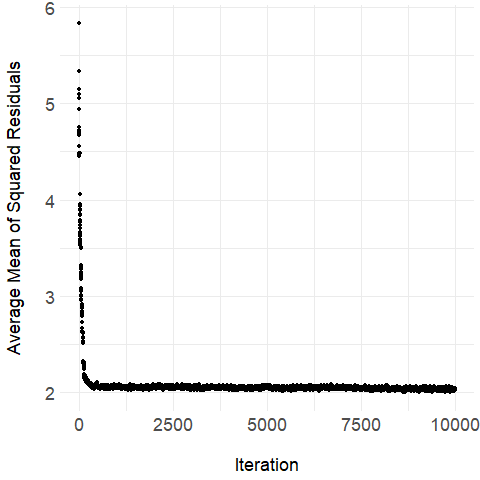}}
    \subfigure[Average of the estimated log spectrum]{\includegraphics[width=0.45\textwidth]{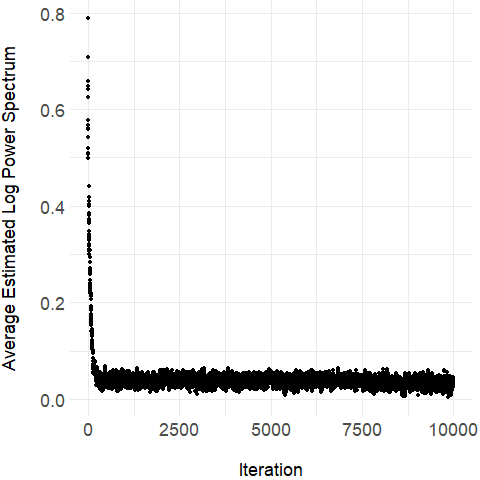}}
    \subfigure[Number of nodes of each tree]{\includegraphics[width=0.45\textwidth]{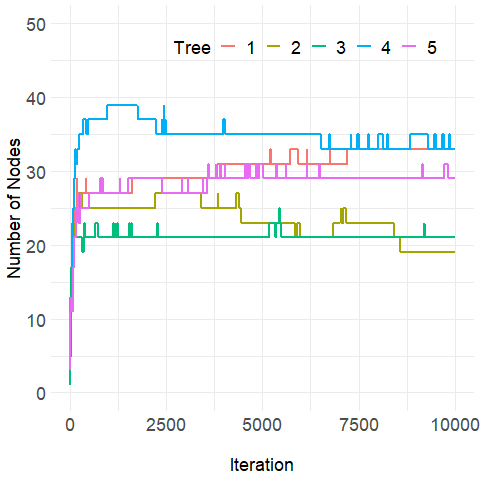}}
    \subfigure[Number of bottom nodes across all trees]{\includegraphics[width=0.45\textwidth]{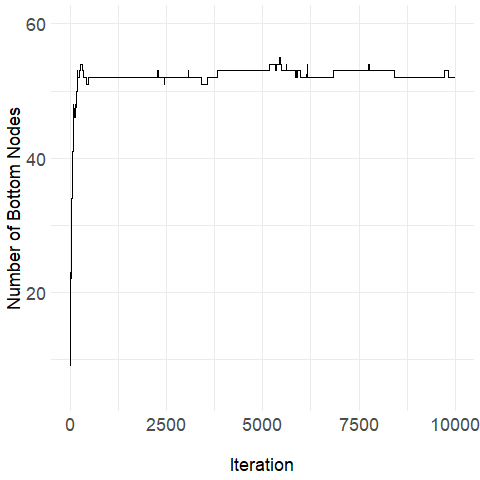}}
   \caption{
    Convergence diagnostics for the Adjusted-AdaptSPEC-X simulation for one replication. Plot (a) contains trace plots of the average mean squared residuals across all time series; Plot (b) shows the average estimated log power spectrum across frequencies and all time series for each iteration; Plot (c) is the trace plots of the total number of nodes for each of the five trees separately (c); Plot (d) is the total number of bottom nodes across all five trees.}
  \label{fig:adaptX_diagnostic}
\end{figure}

 To show the estimation accuracy of the proposed method, for the AR-Friedman simulation setting, where the covariates influence the power spectrum in more complicated ways, We randomly selected eight observations for illustration, and Figure \ref{fig:Est_friedman} shows the estimated log power spectrum of the selected observations. For the Adjusted-AdaptSPEC-X simulation, two observations are randomly selected within each of the four regions shown in Figure \ref{fig:AdaptX_scatter}. The corresponding estimation results for the eight observations are shown in Figure \ref{fig:Est_adaptX}. 
We observe that the proposed model performs very well in capturing the true trend of the power spectrum in both simulation settings. The convergence diagnostic plots for all three settings are shown in Figure \ref{fig:abrupt_diagnostic}, Figure \ref{fig:friedman_diagnostic}, and  \ref{fig:adaptX_diagnostic} separately, which appear to converge after 5000 burn-in iterations.

\begin{figure}
  \centering
    \includegraphics[scale=0.5]{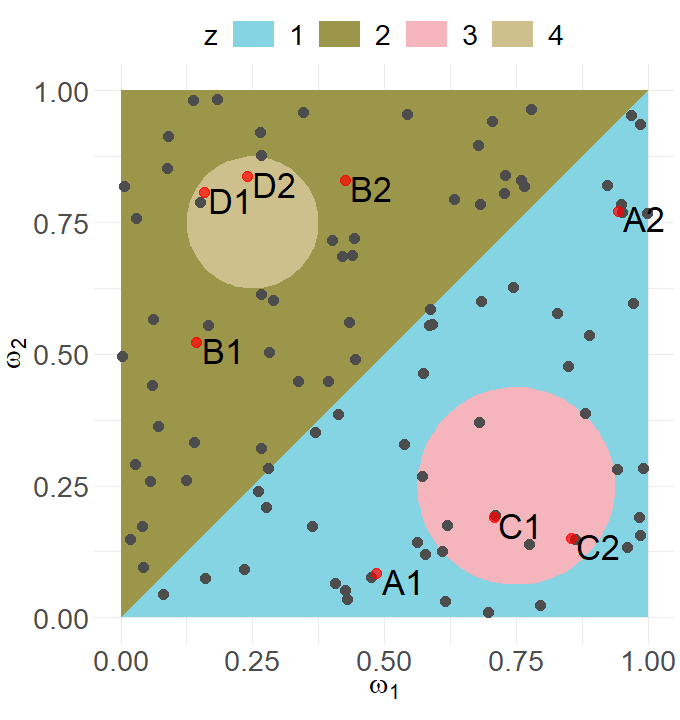}
   \caption{The mapping of the covariates $\omega_1$ and $\omega_2$ to the latent variable $z$ for the Adjusted-AdaptSPEC-X simulation setting with eight specific time series (red dots). }
  \label{fig:AdaptX_scatter}
\end{figure}

\subsection*{Convergence Diagnostics of Gait Maturation}
\begin{figure}
  \centering
    \subfigure[Average mean of squared residuals]{\includegraphics[width=0.45\textwidth]{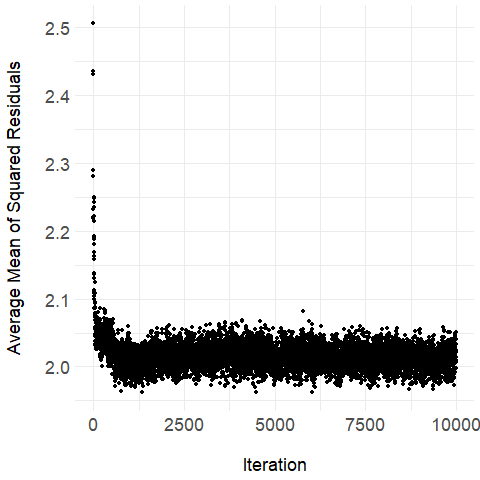}}
    \subfigure[Average of the estimated log spectrum]{\includegraphics[width=0.45\textwidth]{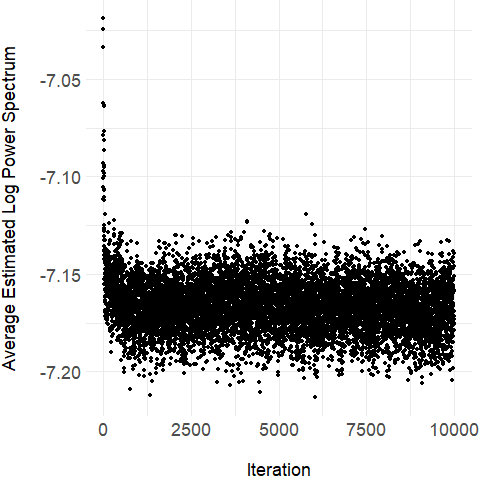}}
    \subfigure[Number of nodes of each tree]{\includegraphics[width=0.45\textwidth]{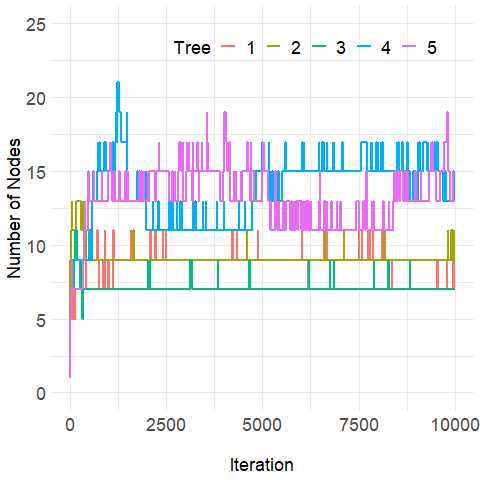}}
    \subfigure[Number of bottom nodes across all trees]{\includegraphics[width=0.45\textwidth]{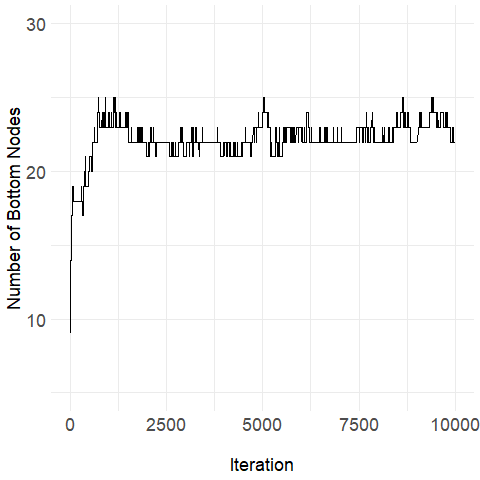}}
   \caption{
    Convergence diagnostics for the gait maturation data analysis.  Plot (a) contains trace plots of the average mean squared residuals across all time series; Plot (b) shows the average estimated log power spectrum across frequencies and all time series for each iteration; Plot (c) is the trace plots of the total number of nodes for each of the five trees separately (c); Plot (d) is the total number of bottom nodes across all five trees.}
  \label{fig:diagnostic_gait}
\end{figure}

Convergence diagnostics for the gait maturation data analysis are shown in Figure \ref{fig:diagnostic_gait}.  While the sampler appears to converge more slowly than the simulation settings, the sampler appears to converge after 5,000 burn-in iterations.

\subsection*{Posterior Probability of Inclusion for Sparse Covariate Effects}
\begin{figure}
    \centering
    \subfigure{\includegraphics[width=0.45\textwidth]{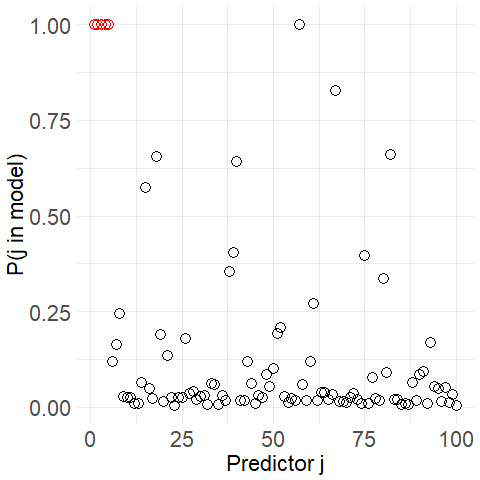}} 
    \subfigure{\includegraphics[width=0.45\textwidth]{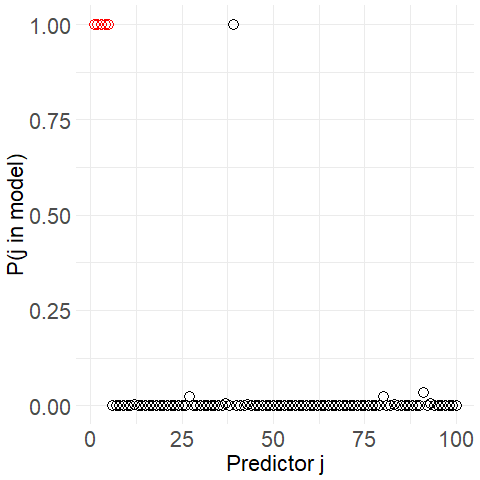}} 
    \caption{Posterior probabilities of model inclusion under the uniform (left) and Dirichlet (right) priors for important variables (red) and noise variables (black).}
    \label{fig:sparsity}
\end{figure}

To demonstrate variable selection using the Dirichlet prior, effects on high-dimensional data as described in Section 5.4, posterior probabilities of model inclusion under the uniform and Dirichlet priors for important variables and noise variables from a single run with $L=500$, $T=250$, and $M=50$ is presented in Figure \ref{fig:sparsity}.

\subsection*{Tree Size and Run Time}
\begin{figure}
  \centering
    \includegraphics[scale=0.45]{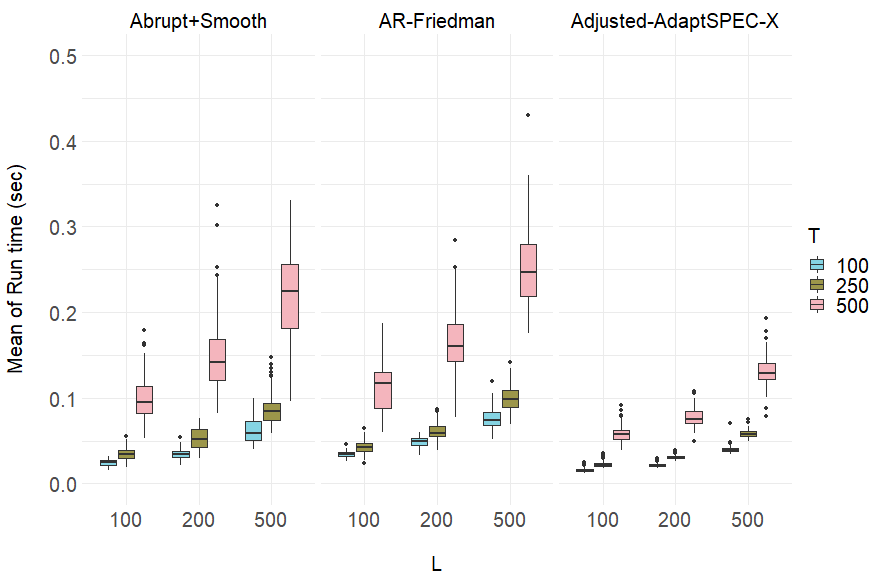}
   \caption{The distribution of mean run times in seconds for a single tree update over 100 replicates of the three simulations with $M=5$ trees.}
  \label{fig:runtime}
\end{figure}

\begin{figure}
  \centering
    \includegraphics[scale=0.55]{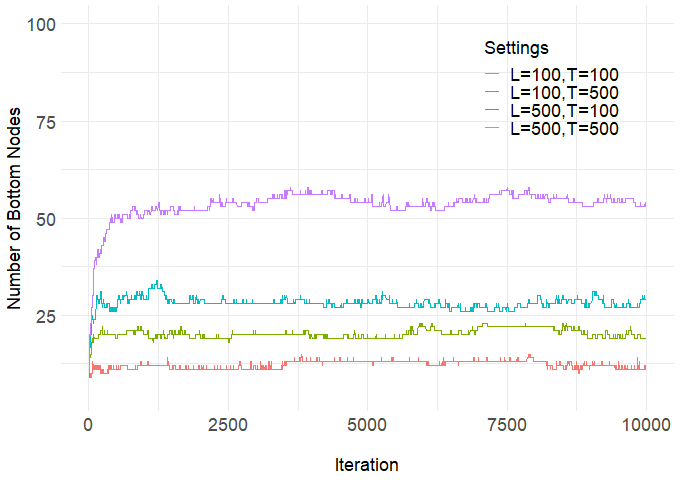}
   \caption{The total number of bottom nodes over iterations for four runs with different $L$ and $T$ for the AR-Friedman simulation setting.}
  \label{fig:contrast_bottom_nodes}
\end{figure}

The distribution of mean run times for the three simulations is shown in Figure \ref{fig:runtime}. We observe that the tree size tends to increase as both the number of time series ($L$) and the length of time series ($T$) increase. Figure \ref{fig:contrast_bottom_nodes} shows the total number of bottom nodes for four different settings for AR-Friedman simulation. The total number of bottom nodes for $L=500, T=500$ is over 50 while it is about 13 for $L=100, T=100$. This can help explain why the mean run times grow more slowly in $L$ compared to $T$, as the mean number of time series belonging to each terminal node increases modestly.

\label{lastpage}

\section*{Web Appendix C: \texttt{R} Implementation of the Proposed Adaptive Bayesian Sum of Trees Model}
\label{sec:C}
BayesSumOfTreesSPEC.zip contains a README file with instructions for running a demo of the Bayesian sum of trees model on the three simulation settings and a description of the files provided.  For the simulation settings, this folder includes the \texttt{R} code needed to generate the simulated data, produce convergence diagnostics, and visualize the estimated covariate-dependent conditional power spectrum.  For the application, this folder includes the original publicly-available data \citep{PhysioNet}, \texttt{R} code for data pre-processing following \cite{HausdorffJ.M1999Mogd}, processed data used for analysis, and \texttt{R} code for producing the convergence diagnostics and ALE plots.  Computationally-intensive aspects of the code, such as the optimization required to approximate the posterior distribution of the spline coefficients and building and modifying the tree structures within the MCMC sampler, are written in C++ using RcppArmadillo \citep{rcpparmadillo} for more efficient computation and reduced run times.

\vspace*{-8pt}

\appendix




\label{lastpage}

\end{document}